\documentclass[useAMS,usenatbib]{mn2e}

\usepackage{natbib}

\usepackage{latexsym,graphicx}
\usepackage{color}
\usepackage{fixltx2e}
\usepackage{verbatim} \usepackage{float}
\usepackage{amsmath,amssymb}
\usepackage{times}
\usepackage{soul}

\usepackage{mathrsfs}
\usepackage{booktabs}

\usepackage{setspace}
\usepackage{rotating}
\usepackage{pdflscape}
\usepackage{subfig}

\usepackage[squaren, Gray, cdot]{SIunits}

\newcommand\hii{H\,{\sc ii} \,}

\def\apgt{\ {\raise-.5ex\hbox{$\buildrel>\over\sim$}}\ }
\def\aplt{\ {\raise-.5ex\hbox{$\buildrel<\over\sim$}}\ }

\let\oldhat\hat
\renewcommand{\hat}[1]{\oldhat{\mathbf{#1}}}

\newcommand{\dm}{\textcolor{black}}
\newcommand{\DM}{\textcolor{black}}
\newcommand{\SN}{\textcolor{black}}

\title[Massive star formation with stellar inertia]{\DM{The burst mode of accretion in massive star formation with stellar inertia}}

\author[D. M.-A.~Meyer et al.]
       {D. M.-A.~Meyer$^{1}$\thanks{E-mail: dmameyer.astro@gmail.com}, 
       E.~I.~Vorobyov$^{2,3}$,
       V.~G.~Elbakyan$^{4}$, 
       S.~Kraus$^{5}$,
       S.-Y.~Liu$^{6}$, \newauthor
       S.~Nayakshin$^{4}$,  
       A.~M.~Sobolev$^{7}$. 
       \\ 
       $^{1}$Institut f\" ur Physik und Astronomie, Universit\" at Potsdam, 
       Karl-Liebknecht-Strasse 24/25, 14476 Potsdam, Germany \\
       $^{2}$Institute of Astronomy, Russian Academy of Sciences, 48 Pyatnitskaya St., Moscow, 119017, Russia \\       
       $^{3}$University of Vienna, Department of Astrophysics, Vienna, 1180, Austria  \\
       $^{4}$School of Physics and Astronomy, University of Leicester, Leicester, LE1 7RH, UK  \\
       $^{5}$University of Exeter, Department of Physics and Astronomy, Exeter, Devon EX4 4QL, UK \\    
       $^{6}$Institute of Astronomy and Astrophysics, Academia Sinica,  
       11F of ASMAB, AS/NTU No.1, Sec. 4, Roosevelt Rd, Taipei 10617, Taiwan \\ 
       $^{7}$Ural Federal University, 19 Mira Str., 620002 Ekaterinburg, Russia \\         
       }

\begin{document}

\date{Received; accepted}

\maketitle

\label{firstpage}

\begin{abstract} 
The burst mode of accretion in massive star formation is a scenario 
linking the initial gravitational collapse of parent pre-stellar cores 
to the properties of their gravitationally unstable discs and of their accretion-driven 
bursts.  
%
%
\DM{In this study,} we present a series of \DM{high-resolution} 3D radiation-hydrodynamics 
numerical simulations for young massive 
stars formed out of collapsing $100\, \rm M_{\odot}$ molecular cores spinning with  
several values of the ratio of rotational-to-gravitational energies $\beta=5\%$--$9\%$. 
\DM{The models include} the indirect gravitational potential caused by disc asymmetries. 
\SN{We find that this modifies the barycenter of the disc, causing significant excursions of the central star position, which we term stellar wobbling.}
%
The stellar wobbling slows down and protracts the development of gravitational 
instability in the disc, reducing the number and magnitude of 
the accretion-driven bursts undergone by the young massive stars, whose properties 
are in good agreement with that of the burst monitored from the massive protostar M17 MIR. 
Including stellar wobbling is therefore important for accurate modeling disc structures. 
Synthetic {\sc ALMA} interferometric images in the millimeter waveband show that 
the outcomes of efficient gravitational instability such as spiral arms and gaseous clumps 
can be detected for as long as the disc is old enough and has already entered the burst 
mode of accretion. 
\end{abstract}

\begin{keywords}
methods: numerical -- radiative transfer -- stars: circumstellar matter. 
\end{keywords}


\section{Introduction}
\label{sect:intro}

The burst mode of accretion in star formation is a model proposed to explain  
\dm{the main accretion phase} of low-mass stars. In this model, gaseous clumps formed 
via disk gravitational fragmentation migrate inwards and trigger accretion bursts when 
tidally destroyed in the vicinity of the star ~\citep{voroboyov_apj_650_2006,vorobyov_apj_719_2010,vorobyov_apj_805_2015,machida_apj_729_2011,
nayakshin_mnras_426_2012,zhao_mnras_473_2018,vorobyov_aa_613_2018}. 
The burst mode of accretion links together several distinct 
phases experienced by young stars, starting from the initial free-fall collapse of 
cold pre-stellar cores, through the main accretion phase when a notable fraction of 
the final stellar mass is acquired during bursts, and finally to the T-Tauri phase, 
in which the characteristics of young stars may still be affected by the earlier burst 
phase \citep{Baraffe2012,vorobyov_aa_605_2017,elbakyan_mnras_484_2019}. 
In this picture, the infall of gas from the collapsing protostellar core plays a key role, replenishing the material lost by the disc during accretion bursts, sustaining the disk in the gravitationally unstable stage and developing disc substructures such as spiral arms and fragments. The burst mode of accretion can resolve the so-called "luminosity problem"~\citep{kenyon_aj_99_1990,kenyon_apj_apj_1990A,kenyon_apjs_101_1995,dunham_apj_747_2012}, 
and the burst characteristics are in accordance with the FU-Orionis-type stars~\citep{vorobyov_apj_805_2015}.

%
 
%
%


A scaling relationship between star-forming mechanisms of low- and high-mass stars has been 
suspected \dm{for decades}~\citep{fuente_aa_366_2001,testi_2003,cesaroni_natur_444_2006,johnston_aa_551_2013}, 
\dm{Observations of} bipolar \hii regions and/or jets in the surroundings of 
young high-mass stars~\citep{Cunningham_apj_692_2009,caratti_aa_573_2015,burns_mnras_467_2017,
burns_2018IAUS,reiter_mnras_470_2017,purser_mnras_475_2018,samal_mnras_477_2018,2019arXiv191208510B,
2019arXiv191111447Z,goddi_apj_905_2020,purser_mnras_504_2021} provided clues and \dm{signatures} 
in its favour. 
This has been strengthened 
by the detection of a first luminous flare from the high-mass protostar 
S255IR-NIRS3~\citep{caratti_nature_2016}, and by many others, 
both in the infrared and by maser emission, in the following 
years~\citep{moscadelli_aa_600_2017,szymczak_aa_617_2018,brogan_apj_866_2018,macleod_mnras_478_2018,
liu_apj_863_2018,hunter_apj_837_2017,provenadzri_mnras_487_2019,lucas_mnras_499_2020,chen_apj_890_2020,
olguin_mnras_498_2020,
chen_natas_2020,burns_natas_2020,2021arXiv210812554C,stecklum_aa_646_2021,hunter_apj_912_2021}.
Additionally, the discovery, in radiation-hydrodynamics numerical simulations, of processes 
for the formation of high-mass stars such as accretion-driven outbursts, similar to that of 
their \dm{low-mass} counterparts, suggested that the burst mode of accretion in star formation also 
applies to higher-mass objects~\citep{meyer_mnras_464_2017}. 
%

It has successively been found that gaseous clumps in accretion discs surrounding young high-mass 
stars have pressure and temperature ($>2000\,\rm K$) consistent with the dissociation of 
\dm{molecular hydrogen}, meaning that disc fragmentation is also at the origin of the formation of 
spectroscopic protobinaries~\citep{meyer_mnras_473_2018}, see also~\citet{oliva_aa_644_2020}. 
The observability of these nascent binaries has been demonstrated by means of 
synthetic {\sc alma} interferometric observations~\citep{meyer_487_MNRAS_2019}. 
Further studies showed that massive young stellar objects \dm{accrete} most of their mass 
within high-magnitude accretion-driven outbursts, taking place during a very short 
fraction of their pre-main-sequence life~\citep{meyer_mnras_482_2019}. 
During the burst \dm{itself}, the accretion rate sharply increases, 
generating a swelling of the protostellar radius, which, in its turn, provokes 
rapid excursions to the cold part of the Herzsprung-Russell diagram and trigger\dm{s} 
the intermittency of the irradiation fluxes \dm{filling} the \hii region of a 
young high-mass star~\citep{meyer_mnras_484_2019}. 
\dm{Lastly}, a parameter study exploring the properties and burst characteristics of a large 
collection of models for the formation of massive young stellar objects concluded 
that no massive protostars should form without bursts, which span over a bimodal 
distribution of short ($\sim 1-10\, \rm yr$) and long ($\sim 10^{3}-10^{4}\, \rm yr$) 
flares~\citep{meyer_mnras_500_2021}, see also~\citet{elbakyan_aa_651_2021}.

Gravitation-hydrodynamics naturally includes the gravitational interaction experienced 
by the substructures developed in the fragmenting accretion discs forming around young 
stars. Consequently, when the density field of an accretion disc 
\dm{is not described by a symmetric pattern} 
because of the presence of \dm{irregular} spiral arms and/or (migrating) gaseous clumps, 
the center of mass of the disc is not anymore coinciding with that of the location of 
the star. This is particularly true in the case of massive discs forming around 
high-mass protostars~\citep{meyer_mnras_473_2018}. The gravitational interaction 
between the disc and the star consequently generates an acceleration and a 
displacement of the protostar as a result of total gravitational force of the 
disc operating onto the young star. 
\dm{While such a} mechanism is intrinsically taken into account \dm{in} grid-based simulations 
in Cartesian coordinates~\citep{krumholz_apj_656_2007,seifried_mnras_417_2011,klassen_apj_823_2016,
rosen_apj_887_2019} 
and Lagrangian models~\citep{springel_araa_48_2010}, it is \dm{often} neglected in grid-based models 
using a spherical, cylindrical or polar coordinate system in which the star is fixed 
at the \dm{coordinate} origin~\citep{voroboyov_apj_650_2006,oliva_aa_644_2020}.

A number of studies on planet and low-mass star formation have not taken into account 
this gravitational backaction of the disc onto the 
star~\citep{tanaka_apj_565_2002,pickett_apj_590_2003,tanaka_apj_602_2004,ou_apj_667_2007, 
boley_apj_656_2007, tsukamoto_mnras_416_2011,fung_apj_790_2014,szulagyi_apj_782_2014, 
zhu_apj_795_2014,lin_mnras_448_2015} while some other \DM{studies} did~\citep{regaly_aa_601_2017}. 
There, the star-disc barycenter and the geometrical grid center are 
\dm{displaced with respect to each other}, in response to the gravitational 
force of the disc mass exerted onto the protostar, and it provokes a motion 
of the central star through the disc. 
A solution to this problem, accounting for the intrinsic singularity present at the 
\dm{coordinate origin of curvelinear coordinate systems}, consists in time-dependently 
displacing the whole disc \dm{in responce to an acceleration} that the star 
\dm{experiences} as a result of disc \dm{gravity force}. 
This so-called non-inertial frame of reference, \dm{which considers} the gravitational 
response to the disc of the stellar motion, has been introduced 
\dm{in~\citet{michael_mnras_406_2010}.}

We hereby continue our numerical investigation of the burst mode of accretion in massive star 
formation, by performing high spatial resolution simulations of radiation-hydrodynamical disc 
models \DM{for several initial rotation rates of the molecular pre-stellar 
core~\citep{meyer_mnras_464_2017}. The numerical setup is 
augmented with the physics of stellar wobbling~\citep{regaly_aa_601_2017}}. 
We measure from our simulations, both 
the physical properties of the accretion discs, as well as the characteristics of the 
accretion-driven outbursts generated by the star-disc 
system~\citep{meyer_mnras_464_2017,meyer_mnras_473_2018,meyer_mnras_500_2021}. 
\DM{We further discuss our results in the context} of the observational appearance of 
the accretion discs, by performing synthetic 
interferometric images as seen by the {\it Atacama Large Millimeter/submillimeter Array} 
({\sc alma}) when operating at band 6 ($1.2\, \rm mm$). 
Beyond the confirmation of the observability of the disc of massive young stellar 
objects~\citep{krumholz_apj_665_2007,jankovic_mnras_482_2019,meyer_487_MNRAS_2019}, 
these synthetic observables reveal how the outcomes of gravitational instability should 
look in realistic discs.

The outline of this study is organised as follows. In Section~\ref{sect:method} we 
present the numerical methods for both the disc and the indirect potential that we 
use to model the circumstellar medium of young massive stars. In Section~\ref{sect:results} 
we show the effects of the young star's motion onto the evolution of its circumstellar 
disc and on the properties of its pre-main-sequence accretion bursts. 
The caveats of our results are discussed in Section~\ref{sect:discussion}, together 
with the effect of stellar motion on interferometric emission. We conclude in 
Section~\ref{sect:conclusion}.


\section{Method}
\label{sect:method}

In this section we present the methodology used to perform our radiation-hydrodynamical 
simulations of accretion discs surrounding massive young stellar objects.

\subsection{Initial conditions}
\label{sect:sub_ic}

The simulations in this study are initialised with a three-dimensional, rotating pre-stellar 
core of external radius $R_{\rm c}$ and radial density profile, 
\begin{equation}
    \rho(r) = \frac{ ( \beta_{\rho} + 3 ) }{ 4\pi  } \frac{ M_{\rm c} }{  R_{\rm c}^{ \beta_{\rho} + 3 } } r^{ \beta_{\rho} },
    \label{eq:density_profile}
\end{equation}
with $r$ the radial coordinate, $M_{\rm c}=100\, \rm M_{\odot}$ the mass of the molecular 
core and $\beta_{\rho}$ a 
negative exponent taken to $\beta_{\rho}=-3/2$. 
The velocity field is set according to the following angular momentum distribution,
\begin{equation}
    \Omega(R) = \DM{\Omega_{0}(\beta)} \Big( \frac{ R }{ r_{0} } \Big)^{ \beta_{\Omega} },
    \label{eq:momentum_distribution1}    
\end{equation}
with $\Omega_{0}(\beta)$ a normalization \DM{factor depending on the adopted 
$\beta$ ratio}, $r_{0}=20\, \rm au$ and $ R = r \sin(\theta)$ the so-called 
cylindrical radius. One can therefore calculate the ratio of kinetic-to-gravitational energy 
of the pre-stellar core, $\beta=E_{\rm rot}/E_{\rm grav}$, and determine its initial toroidal 
rotation profile $v_{\phi}(R)=R\Omega(R)$, while the other components of the velocity are 
set to $v_{\rm r}=v_{\theta}=0$. The pre-stellar core's gravitational energy reads, 
\begin{equation}
    E_{\rm grav} =  \frac{ \beta_{\rho} + 3 }{ 2\beta_{\rho} + 5 }  \frac{G M_{\rm c}^{2}}{R_{\rm c}}, 
   \label{eq:Egrav}    
\end{equation}
and its rotational kinetic energy is as follows, 
\begin{equation}
    E_{\rm rot} = \frac{ ( \beta_{\rho} + 3 )  }{ 4 ( \beta_{\rho} + 2\beta_{\Omega} + 5 )  }
		  \frac{ \DM{\Omega_{0}(\beta)}^{2} M_{\rm c} r_{0}^{ -2\beta_{\Omega} }  }{  R_{\rm c}^{ -2(\beta_{\Omega} + 1 ) }  } 
                  \int_{ 0 }^{ \pi } d\theta \sin( \theta )^{ 3+2\beta_{\Omega} },
   \label{eq:Erot}                  
\end{equation}
respectively. Our models \DM{explore a parameter space} for several initial 
ratio $\beta$ of the pre-stellar core, spanning from $0.05$ to 
$0.09$~\citep{meyer_mnras_500_2021} and we assume 
that the core is in solid-body rotation ($\beta_{\Omega}=0$). 
The list of models is reported in Table~\ref{tab:models}. 
\DM{
For the sake of completeness and comparison, and to emphasize the importance of 
stellar wobbling, the high-resolution simulations are performed with and without 
stellar wobbling.
}

In this study, midplane-symmetric 3D numerical simulations using a a static grid are 
performed in spherical coordinates $(r,\theta,\phi)$ with a grid 
$[r_{\rm in},R_{\rm c}]\times[0,\pi/2]\times[0,2\pi]$ that is spaced as a logarithm in the radial direction 
$r$, a cosine in the direction $\theta$ and uniform in $\phi$. 
It is made of $256\times41\times256$ grid zones discretising the grid, that is taken 
to have $r_{\rm in}=20\, \rm au$ and $R_{\rm c}=0.1\, \rm pc$, respectively. 
The pre-stellar core is initialised with a constant temperature $T_{\rm c}=10\, \rm K$, 
and the molecular material is treated as an ideal gas. 
As in~\citet{meyer_mnras_464_2017} and~\citet{meyer_mnras_473_2018}, we impose outflow 
boundary conditions at the inner and outer part of the radial direction $r$. The mass 
accretion rate onto the protostar $\dot{M}$ is therefore measured as the gas flux crossing 
$r_{\rm in}$. 
The governing equations of these gravito-radiation-hydrodynamics simulations are solved 
using a scheme that is $2^{\rm nd}$ order in space and time, using the {\sc pluto} 
code~\citep{mignone_apj_170_2007,migmone_apjs_198_2012,vaidya_apj_865_2018}, within its 
version augmented for stellar evolution~\citep{hosokawa_apj_691_2009}, radiation 
transport~\citep{kolb_aa_559_2013} and self-gravity. 
Hence, the method incorporates the protostellar radiation releasing photon from the photosphere 
of the young star and irradiating the inner disc region, before being subsequently diffused 
through the disk by flux-limited diffusion treated within the gray 
approximation~\citep{kolb_aa_559_2013}. 
We refer the reader further interested in the numerical method to~\citet{meyer_mnras_500_2021}.

\subsection{Governing equations}
\label{sect:method_equations}

The equations ruling the evolution of the modelled system read, 
\begin{equation}
	   \frac{\partial \rho}{\partial t}  + 
	   \bmath{\nabla}  \cdot (\rho\bmath{v}) =   0,
\label{eq:euler1}
\end{equation}
\begin{equation}
	   \frac{\partial \rho \bmath{v} }{\partial t}  + 
           \bmath{\nabla} \cdot ( \rho  \bmath{v} \otimes \bmath{v})  + 
           \bmath{\nabla}p 			      =   \bmath{f},
\label{eq:euler2}
\end{equation}
\begin{equation}
	  \frac{\partial E }{\partial t}   + 
	  \bmath{\nabla} \cdot ((E+p) \bmath{v})  =	   
	  \bmath{v} \cdot \bmath{f} ,
\label{eq:euler3}
\end{equation}
standing for the conservation of mass, momentum and energy, where $\rho$ is the 
gas density,  $\bmath{v}$ the gas velocity, $p=(\gamma-1)E_{\rm int}$ the thermal 
pressure and $\gamma=5/3$ the adiabatic index. Additionally, the quantity,  
\begin{equation}
	  \bmath{ f } = -\rho \bmath{\nabla} \Phi_{\rm tot} 
			- \lambda \bmath{\nabla} E_{\rm R} 
			- \bmath{\nabla} \cdot \Big( \frac{ \bmath{F_{\star}} }{ c } \Big) \bmath{e}_{\rm r},
\label{eq:f}
\end{equation}
is the force density vector, where $\lambda$ is the flux limiter, and where $E_{\rm R}$, $\bmath{e}_{\rm r}$
$\bmath{F_{\star}}$, $c$ are the thermal radiation energy density, radial unit vector, stellar 
radiation flux and  the speed of light. Lastly, $E  = E_{\rm int} + \rho \bmath{v}^{2}/2$ represents 
the total, internal plus kinetic energy.

Radiation transfer is calculated by solving the equation of radiation transport 
for $E_{\rm R}$, the thermal radiation energy density,  
\begin{equation}
	  \frac{\partial }{\partial t} \Big( \frac{ E_{\rm R} }{ f_{\rm c}  }  \Big)  + 
	  \bmath{\nabla} \cdot \bmath{F}  =	   
	  -\bmath{\nabla} \cdot \bmath{F_{\star}},
\label{eq:rad1}
\end{equation}
with $f_{\rm c}=1/( c_{\rm v} \rho/4 a T^{3} + 1 )$, and where $c_{\rm v}$ is 
the calorific capacity and $a$ is the radiation constant, respectively.  
It is solved in the flux-limited diffusion approach, with $\bmath{F_{\star}}( R_{\star})$ 
the protostellar flux, i.e. \DM{the photospheric irradiation values are estimated 
using both} the interpolated effective temperature $T_{\rm eff}$ and the stellar 
radius $R_{\star}$ from protostellar evolution models~\citep{hosokawa_apj_691_2009}. 
Similarly, the radiation flux reads $\bmath{ F } = -D \bmath{\nabla} E_{\rm R}$,  
where $D$ is the diffusion constant.  
The irradiating stellar flux is set at the inner boundary at a radius $r$ as, 
\begin{equation}
	  \bmath{F_{\star}}(r)  = \bmath{F_{\star}}( R_{\star}) \Big( \frac{ R_{\star} }{ r } \Big)^{2} e^{-\tau(r)},
\label{eq:rad2}
\end{equation}
with $\tau(r)$ the optical depth of the medium, estimated with both constant gas opacity 
and dust opacities from~\citet{laor_apj_402_1993}. 
Self-gravity is included by solving the Poisson equation for the total gravitational 
potential and the stellar gravitational contribution, respectively.

\begin{table*}
	\centering
	\caption{
	List of the simulation models performed in our study. 
	\DM{The table provides the} initial mass of the molecular pre-stellar core $M_{\rm c}$ (in $M_{\odot}$), 
	its initial rotational-to-gravitational energy ratio $\beta$ (in $\%$), the final simulation time 
	$t_{\rm end}$ and the final stellar mass $M_{\star(}t_{\rm end})$ in each simulation models, respectively. 
	The last column indicates whether the simulation includes stellar wobbling. 
	}
	\begin{tabular}{lccccccr}
	\hline
	$\mathrm{Models}$  &  $M_{\rm c}$ $(M_{\odot})$  &  $\mathrm{Grid}\, \mathrm{resolution}$   &  $\beta$ ($\%$)  &  
	$t_{\rm end}$ ($\rm kyr$)             &  $M_{\star}(t_{\rm end})$       & $\mathrm{Wobbling}$ \\ 
	\hline   
	{\rm Run-256-100$\rm M_{\odot}$-5$\%$-wio}  &   $100$    &  $256\times42\times256$  &  $5$    &  40.0  &  26.3   &  no   \\  	
	{\rm Run-256-100$\rm M_{\odot}$-5$\%$-wi}   &   $100$    &  $256\times42\times256$  &  $5$    &  40.0  &  25.9   &  yes  \\
	{\rm Run-256-100$\rm M_{\odot}$-7$\%$-wio}  &   $100$    &  $256\times42\times256$  &  $7$    &  40.0  &  24.3   &  no   \\  
	{\rm Run-256-100$\rm M_{\odot}$-7$\%$-wi}   &   $100$    &  $256\times42\times256$  &  $7$    &  40.0  &  25.7   &  yes  \\ 
	{\rm Run-256-100$\rm M_{\odot}$-9$\%$-wio}  &   $100$    &  $256\times42\times256$  &  $9$    &  32.5  &  16.0   &  no   \\	
	{\rm Run-256-100$\rm M_{\odot}$-9$\%$-wi}   &   $100$    &  $256\times42\times256$  &  $9$    &  32.5  &  18.5   &  yes  \\
        \hline
	\end{tabular}
\label{tab:models}\\
\end{table*}

\subsection{Stellar inertia}
\label{sect:method_disc}

The stellar motion has been implemented as an additional \dm{indirect} potential 
$\Phi_\mathrm{wobbling}$, to which the force $\vec{F}_\mathrm{wobbling}$ \dm{is associated}, 
following the prescription of~\citet{Hirano_2017Science}. It is the opposite of the 
\dm{disc-to-star} gravitational interaction  
$\vec{F}_\mathrm{wobbling} =  \vec{F}_\mathrm{\star/\mathrm{disc}}  = - \vec{F}_\mathrm{disc/\star}$
The force is calculated in spherical 
coordinates, but the respective contributions from each individual grid cells of the 
computational domain are integrated in Cartesian coordinates. 
\dm{
Hence, the force exerted by a given volume element $\delta V$ of the disc reads, 
\begin{equation}
	\delta \vec{F}_\mathrm{disc/\star} = - G M_{\star} \frac{\delta M_\mathrm{disc}(r)}{r^2} \vec{e}_\mathrm{r}, 
	\label{eq:force_int_A}
\end{equation}
and the total force is therefore, 
\begin{equation}
	\vec{F}_\mathrm{disc/\star} = - G M_{\star} \int_\mathrm{disc}  \frac{\rho(r) \delta V}{r^2} \vec{e}_\mathrm{r}, 
	\label{eq:force_int_B}
\end{equation}
with $M_{\star}$ the stellar mass, G the gravitational constant, $\vec{e}_\mathrm{r}$ the radial 
unit verctor and
$\delta M_\mathrm{disc}(r)=\rho(r) \delta V$ the mass contained in 
the corresponding volume element. 
}

\dm{Assuming the star to be fixed in the non-inertial frame of reference, the radial force 
can be expressed as a function of its projections in Cartesian coordinates. 
We obtain, 
\begin{equation}
	\vec{F}_\mathrm{disc/\star} = \sqrt{ (F_\mathrm{disc/\star}^{x})^{2} +  (F_\mathrm{disc/\star}^{y})^{2} + (F_\mathrm{disc/\star}^{z})^{2} }  \vec{e}_\mathrm{r}, 
	\label{eq:force_cartesian_1}
\end{equation}
with, 
\begin{equation}
	\vec{F}_\mathrm{disc/\star}^{x} = - G M_{\star} \Big[ \int_\mathrm{disc}  \frac{\rho \delta V}{r^2}  \Big]   \vec{e}_\mathrm{r}  \cdot \vec{e}_\mathrm{x}, 
	\label{eq:force_int_C1}
\end{equation}
\begin{equation}
	\vec{F}_\mathrm{disc/\star}^{y} = - G M_{\star} \Big[ \int_\mathrm{disc}  \frac{\rho \delta V}{r^2}  \Big]  \vec{e}_\mathrm{r} \cdot \vec{e}_\mathrm{y}, 
	\label{eq:force_int_C2}
\end{equation}
and 
\begin{equation}
	\vec{F}_\mathrm{disc/\star}^{z} = - G M_{\star} \Big[ \int_\mathrm{disc}  \frac{\rho \delta V}{r^2} 
	\Big] \vec{e}_\mathrm{r}  \cdot \vec{e}_\mathrm{z}, 
	\label{eq:force_int_C3}
\end{equation}
with $\vec{e}_\mathrm{x}$, $\vec{e}_\mathrm{y}$, $\vec{e}_\mathrm{z}$ the Cartesian 
unit vectors. In the disc midplane, 
\begin{equation}
	 \vec{e}_\mathrm{r}  \cdot \vec{e}_\mathrm{x} = \mathscr{C}_{\phi} \mathscr{S}_{\theta}, 
	\label{eq:delta_force_disc1}
\end{equation}
\begin{equation}
	\vec{e}_\mathrm{r} \cdot \vec{e}_\mathrm{y} = \mathscr{S}_{\phi} \mathscr{S}_{\theta}, 
	\label{eq:delta_force_disc2}
\end{equation}
\begin{equation}
	\vec{e}_\mathrm{r}  \cdot \vec{e}_\mathrm{z} =  \mathscr{C}_{\theta}, 
	\label{eq:delta_force_disc3}
\end{equation}
respectively, with $\phi$ and $\theta$ the azimuthal and polar angles of the spherical 
coordinate system, and where $\mathscr{C}_{\theta}=\cos(\theta)$, $\mathscr{S}_{\theta}=\sin(\theta)$, 
$\mathscr{C}_{\phi}=\cos(\phi)$ and $\mathscr{S}_{\phi}=\sin(\phi)$ respectively. 
}

Finally, the effect of the \dm{indirect} force is implemented as an acceleration term, 
\begin{equation}
	\vec{g}^{'} = \frac{ \vec{F}_\mathrm{wobbling} }{ M_{\star}} 
	            = -\frac{ \vec{F}_\mathrm{disc/\star} }{ M_{\star}},  
	\label{eq:acceleration}
\end{equation}
into the solver for gravitation of the code {\sc pluto}. 
Explicitly, the components of $\vec{g}^{'}$ read,
\begin{equation} \begin{split}
	g^{'}_{r}            = \frac{ 1 }{ M_{\star}} \Big[   F_\mathrm{disc/\star} \mathscr{S}_{\theta} \mathscr{C}_{\phi}  +  F_\mathrm{disc/\star} \mathscr{S}_{\theta} \mathscr{S}_{\phi}  +  F_\mathrm{disc/\star} \mathscr{C}_{\theta}  \Big], 
	\label{eq:acc_disc1}
\end{split} \end{equation} 
\begin{equation} \begin{split}
	\vec{g}^{'}_{\phi}   = \frac{ 1 }{ M_{\star}} \Big[   F_\mathrm{disc/\star} \mathscr{C}_{\theta} \mathscr{C}_{\phi}  +  F_\mathrm{disc/\star} \mathscr{C}_{\theta} \mathscr{S}_{\phi}  -  F_\mathrm{disc/\star} \mathscr{S}_{\theta}  \Big], 
	\label{eq:acc_disc2}
\end{split} \end{equation}
\begin{equation}
	\vec{g}^{'}_{\theta} = \frac{ 1 }{ M_{\star}} \Big[   -F_\mathrm{disc/\star} \mathscr{S}_{\phi}  +  F_\mathrm{disc/\star} \mathscr{C}_{\phi}   \Big],  
	\label{eq:acc_disc3}
\end{equation}
respectively. 
The stellar mass $M_{\star}$ is evaluated at each timesteps, as the time integral of the accretion rate 
$\dot{M}$ onto the sink cell.

\begin{figure*}
        \centering
        \includegraphics[width=0.8\textwidth]{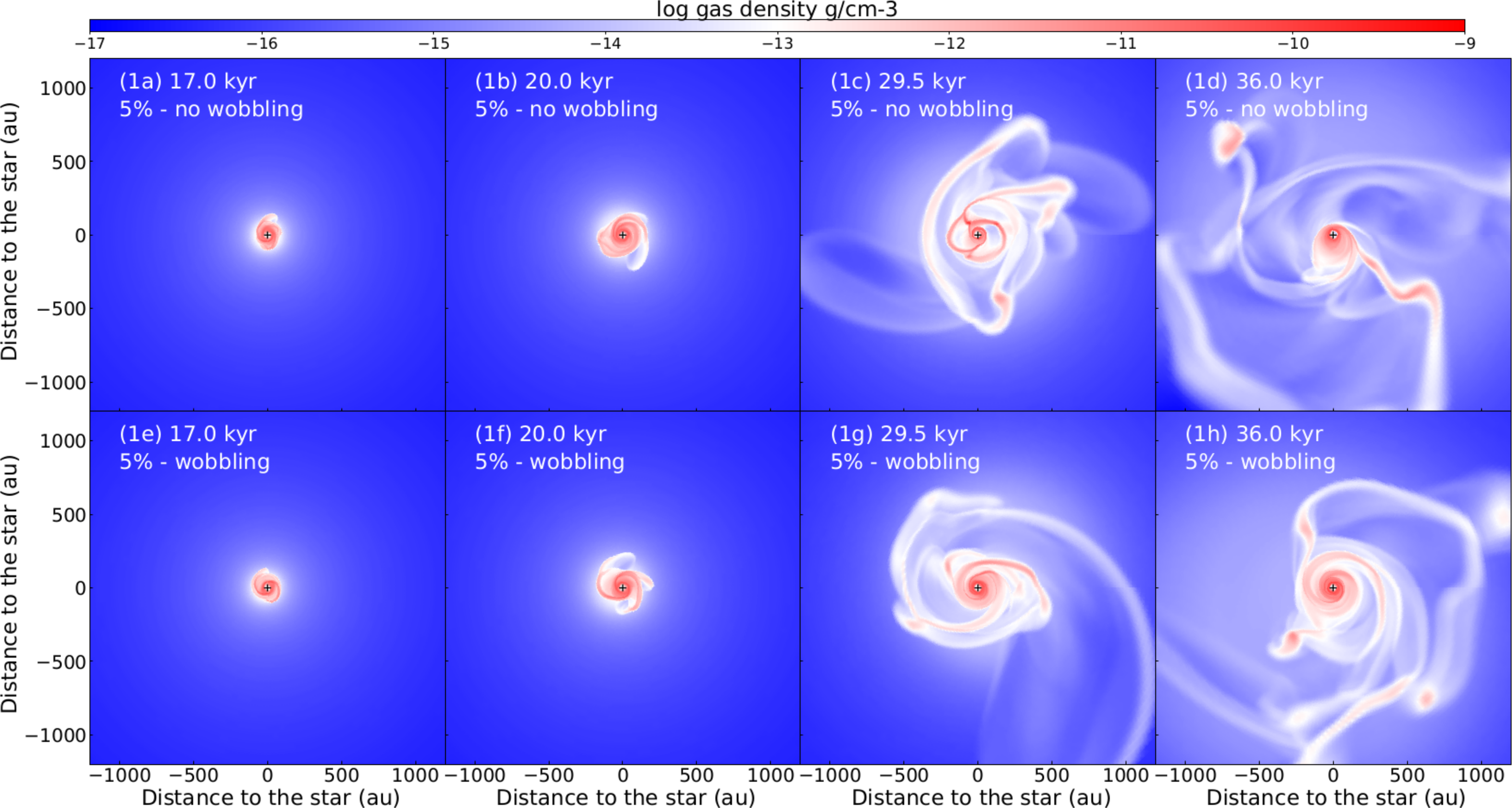} \\     
        \includegraphics[width=0.8\textwidth]{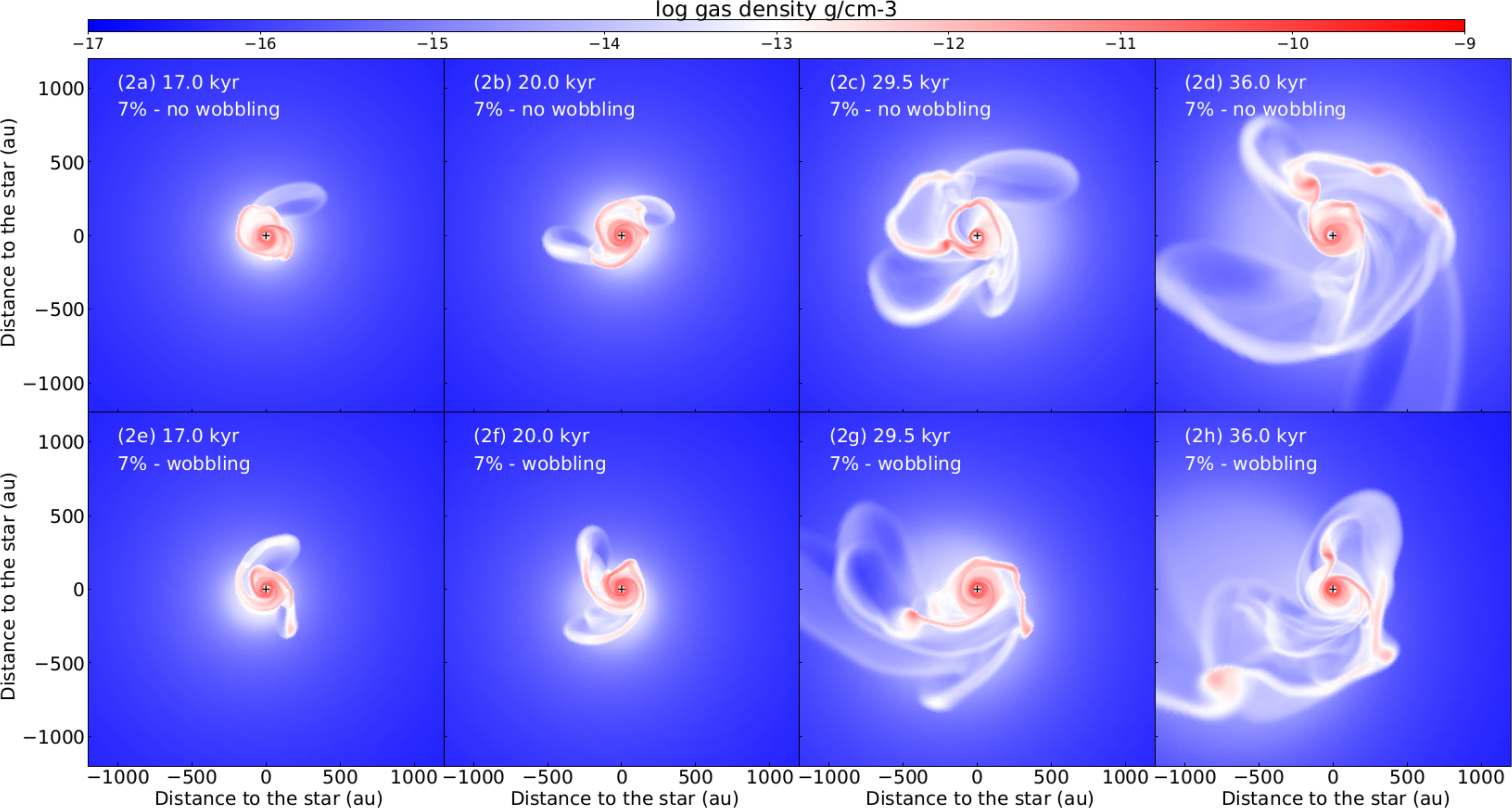} \\
        \includegraphics[width=0.8\textwidth]{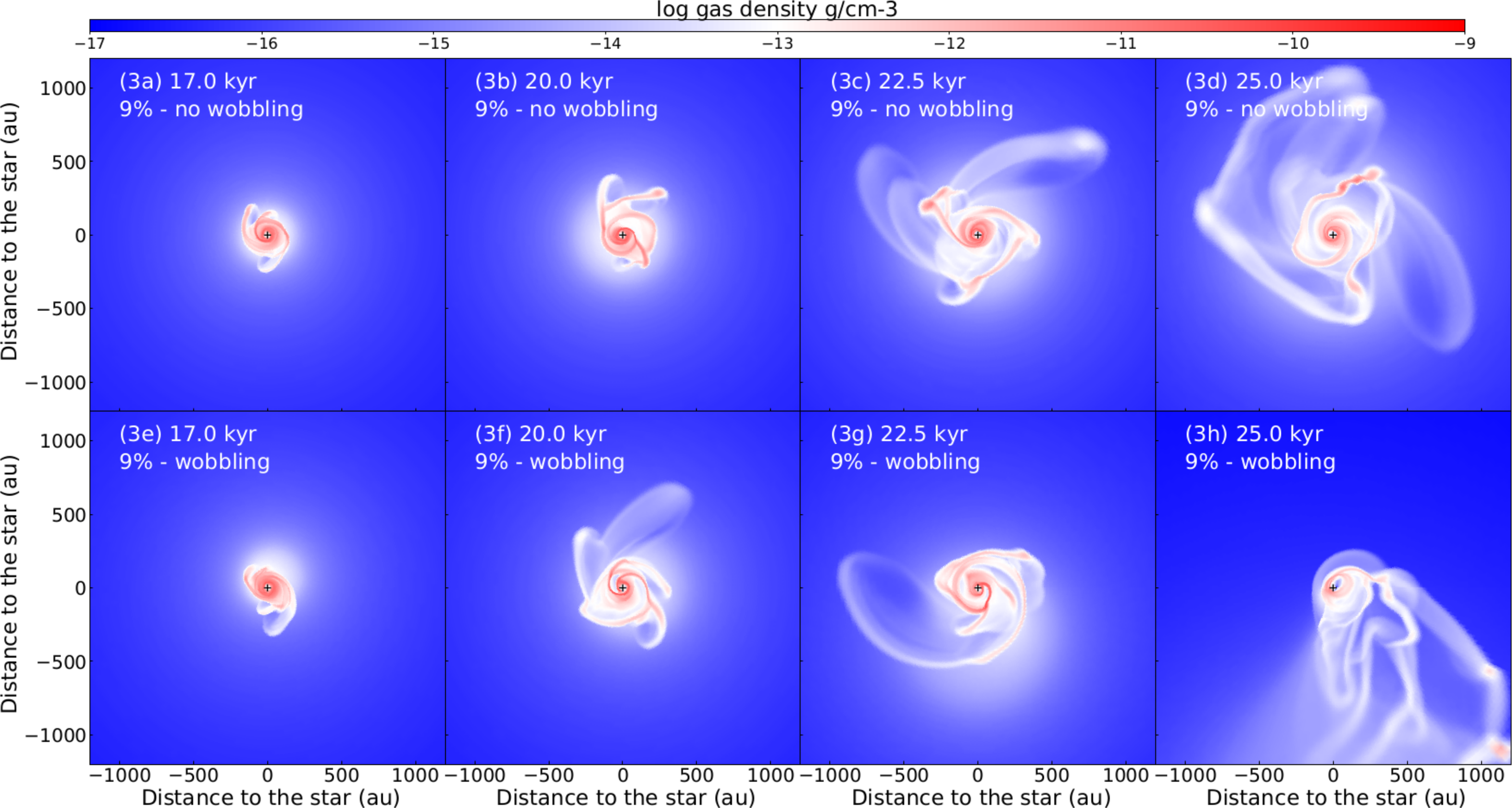}         
        \caption{
        Density fields in the disc simulations with $\beta=5\%$ (top), $\beta=7\%$ 
        (middle) and $\beta=9\%$ (bottom). The figures display the disc \textcolor{black}{midplane 
        gas density (in $\rm g\, \rm cm^{-3}$)}, for several characteristics time instances 
        of the accretion disc evolution, without (upper panels) and with (lower panels) 
        stellar wobbling. 
        }
        \label{fig:discs}  
\end{figure*}


\section{Results}
\label{sect:results}

This section explores the evolution of the protostellar disc properties such as the disc radius  
and mass, for both simulations without and with the inclusion of the stellar motion of the 
central high-mass star. Similarly, we present how the intensity and duration of the accretion-driven 
bursts change when the star is allowed to move, for several initial conditions of the pre-stellar core.

\begin{figure*}
        \centering
        \includegraphics[width=0.95\textwidth]{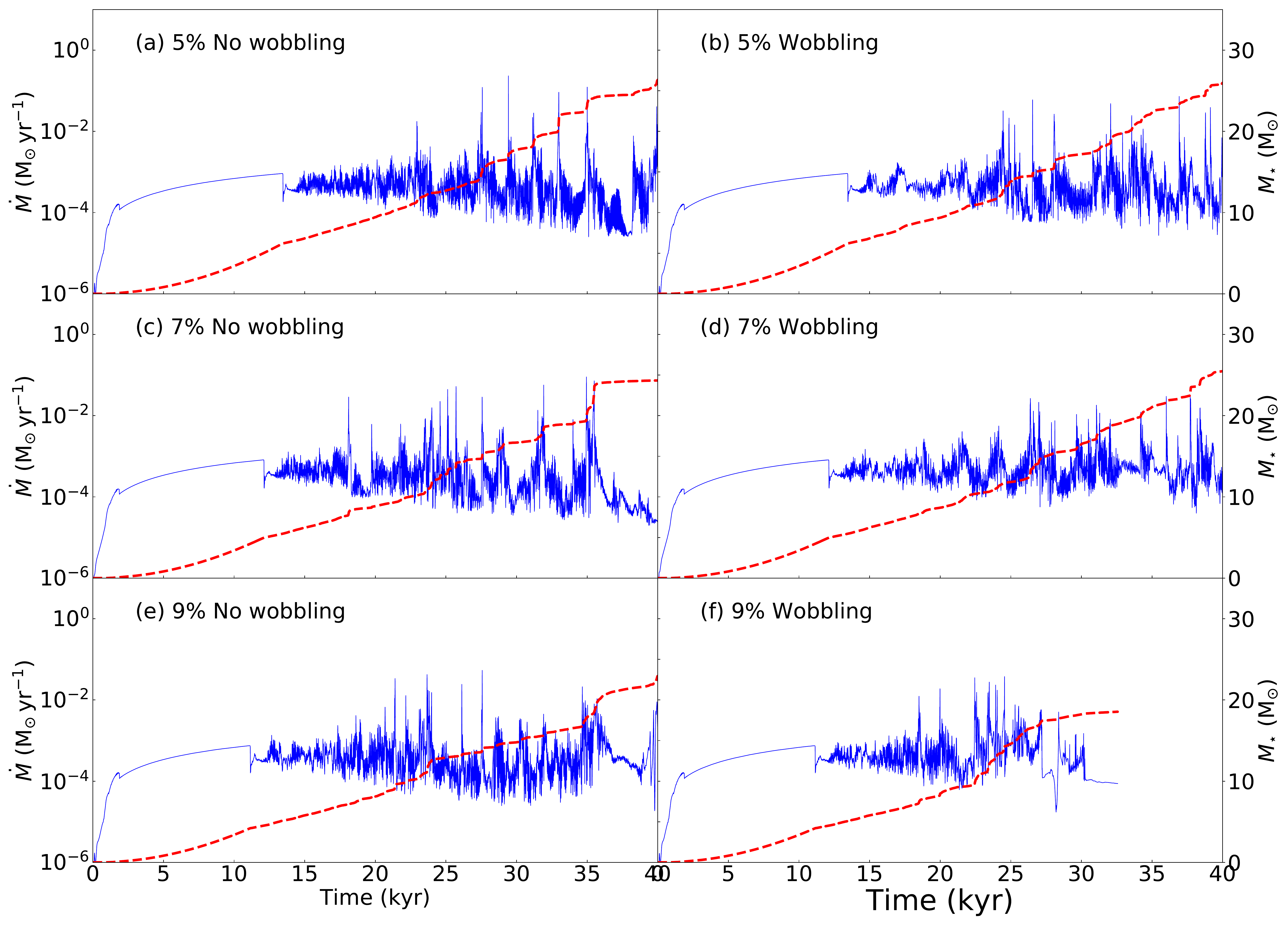} 
        \caption{
        Accretion rate onto the \dm{protostar} (in $\rm M_{\odot}\, \rm yr^{-1}$, 
        solid blue line) and evolution of the mass of the young massive stellar objects 
        (in $\rm M_{\odot}$) for our simulation models. 
        The simulations are displayed for several $\beta$-ratio spanning from $5\%$ 
        (top) to $9\%$ (bottom), without (left) and with (right) stellar motion. 
        }
        \label{fig:accretion_rate}  
\end{figure*}

\subsection{Discs}
\label{sect:sub_discs}

\subsubsection{Structure and evolution}
\label{sect:disc_evolution}

In Fig.~\ref{fig:discs} we plot the midplane density field of the accretion discs 
(in $\rm cm^{-3}$) in our simulations with initial kinetic-to-gravitational 
energy $\beta=5\%$ (top series of panels), $\beta=7\%$ (middle series of panels) 
and $\beta=9\%$ (bottom series of panels). For each model, the discs are shown at 
\dm{several time instances} as well as a later time, both without and with 
stellar inertia.  
In the figures, we only represent the inner ($\le 1000\, \rm au$) part of the computational 
domain, for time instances \dm{that are} older than the end of the free-fall gravitational collapse of 
the pre-stellar cloud and the onset of the disc formation. 
The accretion discs without \dm{stellar wobbling} qualitatively evolve as described in the 
previous papers of this series, by undergoing efficient gravitational instabilities 
responsible for the formation of spiral arms and other rotating, self-gravitating 
substructures such as gaseous clumps and disc 
fragments~\citep{meyer_mnras_473_2018,meyer_487_MNRAS_2019}. 
Notable differences in the models with stellar inertia compared to the simulations 
with fixed star appear at times $\ge 25\, \rm kyr$.

The models without and with stellar inertia initially do not reveal differences, regardless \DM{of} 
the $\beta$-ratio of the collapsing molecular cloud. This is visible in the left 
column of each panels showing the discs at time $17\, \rm kyr$. Their size increases 
as a function of the increasing $\beta$-ratio of the molecular cloud, because the 
timescale of the free-fall gravitational collapse is shorter for high $\beta$-ratio 
(see Fig.~\ref{fig:accretion_rate}), hence at time $17\, \rm kyr$ the discs with 
$\beta=9\%$ are slightly older and more developed than that with $\beta=5\%$. 
The same is true at time $20.0\, \rm kyr$. Even tough the discs are qualitatively 
similar in terms of overall structure, differences begin to appear in their morphology, 
such as the number and the geometry of the growing spiral arms, see 
Fig.~\ref{fig:discs}-1b,1f.

In our model with $\beta=5\%$, one can see at time $29.5\, \rm kyr$, that the disc 
with \dm{stellar} wobbling has a rather compact structure wrapped by a large spiral arm 
(Fig.~\ref{fig:discs}-1g), while the simulation without stellar wobbling  
\dm{has a more complex inner structure} and exhibit signs of fragmentation both at its 
center and in a spiral arm extending to distance $>500\, \rm au$ from the protostar. 
The parent arm hosts a dense clump in the process of detaching from it, before 
migrating down to the protostar (Fig.~\ref{fig:discs}-1c). 
At time $36.0\, \rm kyr$, both discs exhibit signs of active gravitational instability. 
The disc model without wobbling is more extended and has a (northern) very 
massive and large clump $\sim 1000\, \rm au$ from the protostar as 
well as a (southern) dense spiral arm (Fig.~\ref{fig:discs}-1d), 
while the model with wobbling has a rounder and more compact structure 
in which enrolled spiral arms host a few clumps at 
distances $<500\, \rm au$ from the protostar (Fig.~\ref{fig:discs}-1h). 
The discs are at different stages of internal reorganisation after it has experienced 
migrations of dense clumps onto the protostar, which \dm{generates an ejection of 
the clumps' gaseous tails} that reshuffle the whole 
disc structure~\citep{meyer_mnras_473_2018}.  

\DM{
Hence, stellar wobbling moderately delays the fragmentation of the disc and 
reduces its entire size. 
The explanation is that when stellar wobbling is not considered, the accretion disc alone 
\textcolor{black}{contains} all the angular momentum of the star-disc system. Permitting 
the protostar to move 
(i) redistributes this angular momentum between the star and the disk with a net effect that the disk angular momentum decreases. Without stellar 
inertia, the discs turn to be of larger radius, and, therefore, their more extended 
spiral arms are more prone to fragment because disk fragmentation occurs predomenantly 
at large radii \citep{johnson_apj_597_2003}.
}

The above described effects of stellar wobbling in the disc evolution is equivalently 
found in the other simulation models with $\beta=7\%$ and $\beta=9\%$. The density 
field in the model with $\beta=7\%$ indeed displays similar features, with a disc 
with stellar inertia that fragments quicker than in the model with $\beta=5\%$, see 
Fig.~\ref{fig:discs}-2c and Fig.~\ref{fig:discs}-2g. 
Again, stellar inertia seems to make the circumstellar 
environment of the protostar more compact, with a smaller number of spiral arms. 
At time $36.0\, \rm kyr$ the disc with $\beta=7\%$ without \dm{stellar} wobbling carries a 
larger number of clumps than the model with \dm{stellar} wobbling, although they exhibit a 
similar qualitative level of fragmentation. 
Lastly, \DM{the} model with $\beta=9\%$ slightly deviates from the simulations with 
$\beta\le7\%$. Because of its faster initial rotational velocity of the gas 
in the pre-stellar core, the development of gravitational instability in the disc 
\dm{takes more time than in the other models, see Fig.~\ref{fig:discs}-3c and 
Fig.~\ref{fig:discs}-3f.  
}
At later time ($36.0\, \rm kyr$), the effects of the \dm{indirect} potential accounting 
for stellar inertia becomes so strong that the disc is {strongly} destroyed 
(see discussion on boundary effects in Section~\ref{sect:discussion_caveats}).

\begin{figure}
        \centering
        \includegraphics[width=0.49\textwidth]{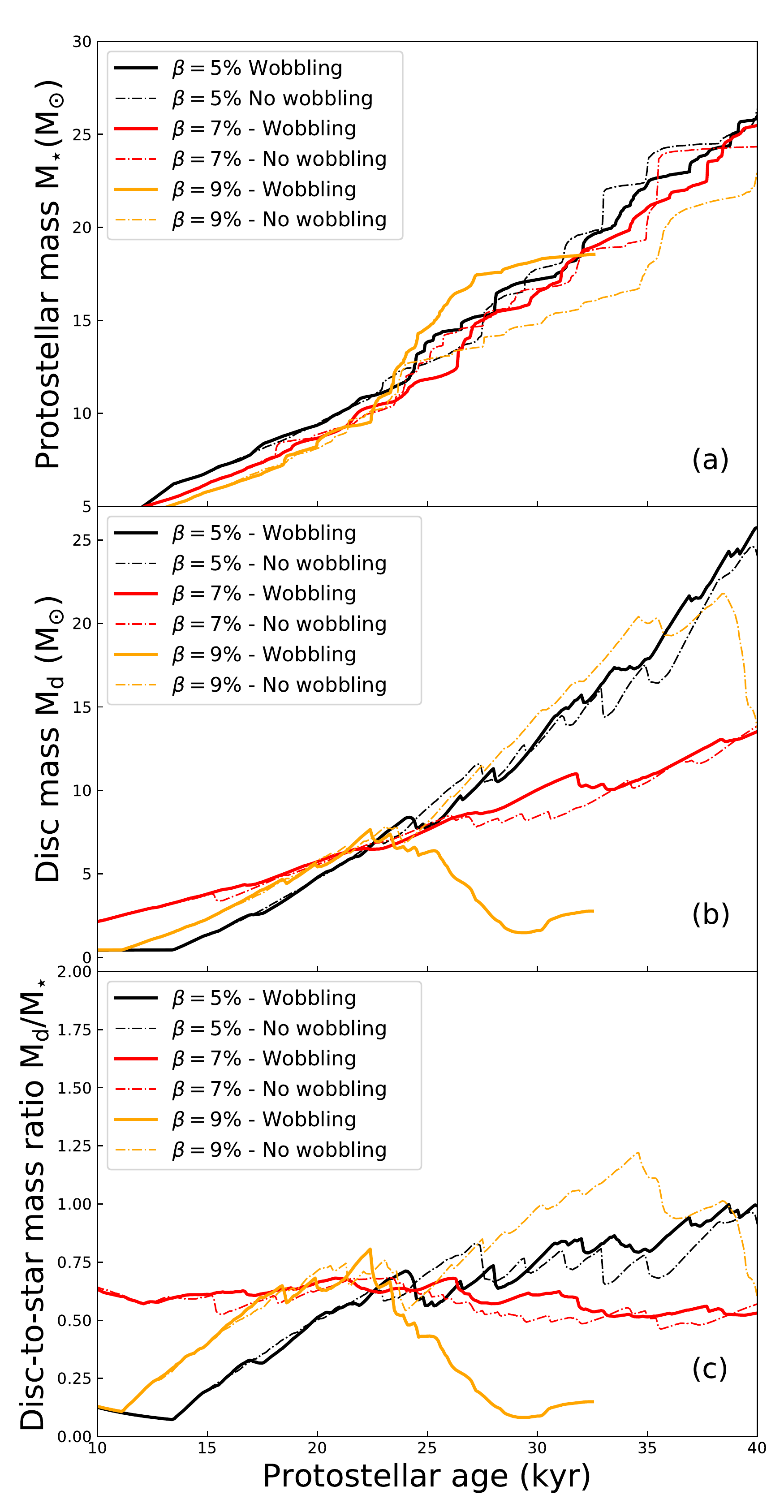} 
        \caption{
        Protostellar mass (a), \textcolor{black}{disc mass (b) and disc-to-star mass ratio (c)} in our 
        disc models with $\beta=5\%$ (black), $\beta=7\%$ (red) and $\beta=9\%$ (orange). 
        Data for both models with (solid lines) and without (dashed-dotted lines) stellar 
        wobbling are shown in the figure. 
        }
        \label{fig:mass_disc}  
\end{figure}

\begin{figure*}
        \centering
        \includegraphics[width=0.7\textwidth]{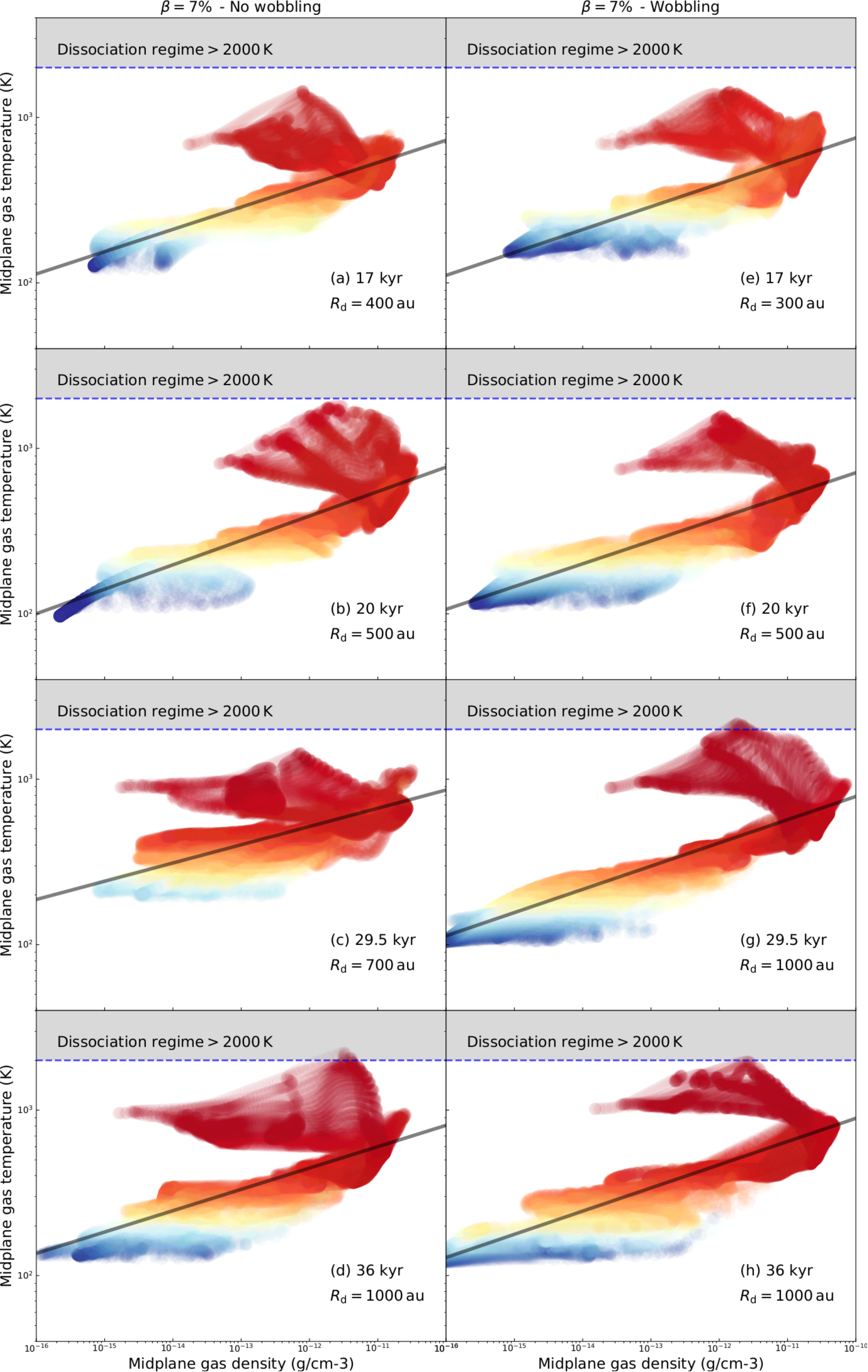} 
        \includegraphics[width=0.7\textwidth]{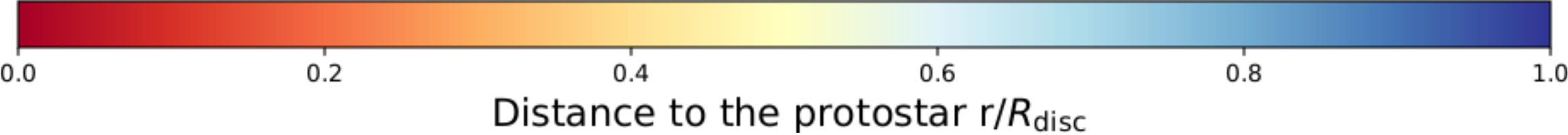} 
        \caption{
        Evolution of the disc midplane in the density-temperature diagram for 
        the model with $\beta=7\%$, without (left panels a-d) and with stellar 
        inertia (right panels e-h), for different time from $17\, \rm kyr$
        to $36\, \rm kyr$ after the beginning of the simulation. 
        The color indicate the distance of each midplane gas element to the 
        central protostar, normalised to the size of the disc at the considered 
        time instance. 
        The gray zone highlight the region with $T\ge2000\, \rm K$ in which the 
        gas is hotter than the molecular dissociation temperature. 
        The black line is a fit to the entire dataset. 
        }
        \label{fig:diagram}  
\end{figure*}

\subsubsection{Protostellar accretion rate}
\label{sect:accretion_rates}

Fig.~\ref{fig:accretion_rate} shows the accretion rate histories (thin solid blue line, 
in $\rm M_{\odot}\, \rm yr^{-1}$) and the stellar mass (thick dashed red line, 
in $\rm M_{\odot}$) of the young massive stellar objects forming in our simulations with 
kinetic-to-rotational energy radio $\beta=5\%$ (a,b), $7\%$ (c,d), $9\%$ (e,f). 
The left panels show the accretion rates of the models without wobbling, while the 
right panels plot the models with wobbling, respectively. 
In the panels, one can see the initial increase of the accretion 
rate during the free-fall collapse of the pre-stellar material, up to the time of the 
onset of the disc formation, at time $\approx 12\, \rm kyr$, when the accretion rate 
becomes variable because of azimuthal anisotropies in the accretion flow. 
At this time, the disc is small and of moderate size (Fig.~\ref{fig:discs}-1a) and 
\dm{can be described to be } in the so-called quiescent mode of accretion~\citep{meyer_mnras_482_2019}.

At times $\approx 26\, \rm kyr$, \DM{the accretion} rate history displays several strong 
peaks corresponding to the \dm{transport} of disc material through the sink cell at rates 
$\ge 10^{-2}\, \rm M_{\odot}\, \rm yr^{-1}$. 
Those features characterise the burst mode of accretion in which the protostar 
enters when accreting dense segments of spiral arms and/or migrating gaseous clumps, 
and, this translates into luminous accretion-driven 
flares~\citep{meyer_mnras_464_2017,meyer_mnras_482_2019}. 
\dm{Star-disc systems in the burst mode are} associated with large discs of complex internal 
structures (Fig.~\ref{fig:accretion_rate}b).  
The rest of the modelled disc evolution is constituted of a succession of episodic 
accretion bursts interspersed by quiescent phases of accretion, which is characteristic 
of young stars growing and gaining mass within the burst mode of 
accretion~\citep{meyer_mnras_482_2019,meyer_mnras_500_2021}. 
The evolution of the protostellar mass reflects this by displaying \dm{a series of step-like 
increases} at the moment of the bursts (dotted red line of Fig.~\ref{fig:accretion_rate}a), 
see also~\citet{meyer_mnras_473_2018}.

\textcolor{black}{
The accretion rate history of the model with $\beta=5\%$ including wobbling 
does not display strong qualitative differences in terms of accretion bursts 
compare to the models without} 
stellar inertia, since the disc fragments (Fig.~\ref{fig:discs}-1c,1g) 
and the protostar enters the burst mode of accretion at similar times ($\approx 24\, \rm Myr$). 
\dm{The same is true in our simulation with $\beta=7\%$.} 
The accretion rate of the model with stellar wobbling and $\beta=9\%$ is much more erratic 
as a consequence of the dramatic distortion of the disc by the stellar wobbling, 
see Fig.~\ref{fig:discs}-3h. 
Because the accretion bursts of the models with wobbling are of reduced magnitude, 
the mass evolution \dm{does} not exhibit the large step-like increase of the mass 
evolution in the models with wobbling. 
\dm{
In all our models, the magnitude and the number of the strongest accretion events 
is reduced when wobbling is included, see our Table~\ref{tab:A}. 
}

\subsubsection{Disc size, mass and disc-star mass ratio}
\label{sect:size}

Fig.~\ref{fig:mass_disc} plots the properties of the simulated star-disc systems, such as the protostellar 
mass (top panel, in $\rm M_{\odot}$), the disc mass (middle panel, in $\rm M_{\odot}$) 
and the disc-to-star mass ratio (bottom panel). On each panel, the quantities are shown 
for all 6 models of Table~\ref{tab:models} \DM{represented in the figures} by a color coding, 
with thick solid lines (models without wobbling) and thin dashed lines (models with 
wobbling) for the simulations with $\beta=5\%$ (black), $\beta=7\%$ (red), $\beta=9\%$ 
(yellow). 
One can see that the protostellar mass histories are smoother in the simulations with 
wobbling than in those without, since, as discussed above, the onset of the gravitational 
instability happens sooner if no stellar inertia is included. 
\dm{
However, in the models with $\beta=5\%$ and $\beta=7\%$, for the timescales 
of $40\, \rm kyr$ that we simulate, the final masses are similar}, regardless 
of the physics taken into account in the simulation, as those simulations 
end with a final massive young protostellar object of 
$\approx 26\, \rm M_{\odot}$ and $\approx 25\, \rm M_{\odot}$, respectively. 
Inversely, for high $\beta$-ratio of $9\%$, it reveals a notable 
difference of a few solar mass between the stellar mass calculated without 
and with wobbling at times $\approx 32.5\, \rm kyr$, as a direct consequence 
of gravitational instability boundary effects \dm{influencing} the disc evolution in 
this fast-spinning system.

The \dm{middle} panel of the figure (Fig.~\ref{fig:mass_disc}b) shows the evolution of 
the disc mass, calculated by integrating the gas density in a cylinder of dimensions 
$2000\, \rm au$ and height $500\, \rm au$ as in~\citet{klassen_apj_823_2016,meyer_mnras_473_2018}. 
The disc mass evolution $M_{\rm d}$ (in $M_{\odot}$) is marked by abrupt decreases 
corresponding to the \dm{time instances when} dense gaseous clumps disappear 
into the central sink cell of radius $r_{\rm in}=20\, \rm au$, also provoking  
step-like increase of the protostellar mass history (Fig.~\ref{fig:mass_disc}a). 
Indeed, when a clump is accreted and the mass and disc evolution have opposite 
variability, the first gains the amount of mass the latter loses. 
\textcolor{black}{Note that the disc masses} do not sensibly change whether wobbling is \dm{included} 
or not, at least for the time that we simulate ($40\, \rm kyr$), 
\textcolor{black}{except for the model with $\beta=9\%$ that is affected by 
strong wobbling-related boundary issues (Section~4.2).} 
The bottom panel shows the evolution of the disc-to-star mass ratio 
$M_{\rm d}/M_{\star}$ (Fig.~\ref{fig:mass_disc}c) which remains 
$\ge 0.5$ once the accretion discs form~\citep{meyer_mnras_500_2021}, 
throughout all the simulations,  except for the model with $\beta=9\%$ 
and unfixed star, which deviates from this trend starting 
from time $\approx 25\, \rm kyr$ when the effects of the stellar wobbling are so strong 
that they induce a stretching of the circumstellar disc (see Fig.~\ref{fig:discs}-3c,
Fig.~\ref{fig:discs}-3h and also Section~\ref{sect:discussion_caveats}).  

\begin{table*}
	\centering
	\caption{
	Table displaying the main properties of the accretion bursts in our simulations 
	series with $\beta=5\%$$-$$9\%$. 
	$N_{\mathrm{bst}}$ is the number of bursts undergone by the protostars at a given 
	magnitude class. $L_{\mathrm{max}}/L_{\mathrm{min}}/L_{\mathrm{mean}}$ stands for the 
	maximum, minimum, and mean burst luminosities. Equivalently, $\dot{M}_{\mathrm{max}}/\dot{M}_{\mathrm{min}} 
	/\dot{M}_{\mathrm{mean}}$ correspond to the accretion rates onto the protostar, and $t_{\mathrm{bst}}^{\mathrm{max}}$/$t_{\mathrm{bst}}^{\mathrm{min}}$/$t_{\mathrm{bst}}^{\mathrm{mean}}$
	relate to the accretion burst duration. $t_{\mathrm{bst}}^{\mathrm{tot}}$ is the total time a 
	massive young stellar object spends bursting via a given burst magnitude class. 
	}
        \begin{tabular}{lcccccr}
        \hline  
        Model & $N_{\mathrm{bst}}$  & $L_{\mathrm{max}}/L_{\mathrm{min}}/L_{\mathrm{mean}}$   ($10^{5}\, \rm L_{\odot}$) & $\dot{M}_{\mathrm{max}}/\dot{M}_{\mathrm{min}}/\dot{M}_{\mathrm{mean}}$ ($\rm M_{\odot}\, \rm yr^{-1}$) & $t_{\mathrm{bst}}^{\mathrm{max}}$/$t_{\mathrm{bst}}^{\mathrm{min}}$/$t_{\mathrm{bst}}^{\mathrm{mean}}$ $\rm (yr)$  & $t_{\mathrm{bst}}^{\mathrm{tot}}$ $\rm (yr)$\tabularnewline
        \hline
        \midrule 
        \multicolumn{6}{c}{\textbf{1-mag cutoff}}\tabularnewline
	\midrule 
{\rm Run-256-100$\rm M_{\odot}$-5$\%$-wio}   &   6  &  19.89 / 0.464 / 5.40 & 0.0397 / 0.0073 / 0.0170 & 15 / 4 /  9 &  51\tabularnewline
{\rm Run-256-100$\rm M_{\odot}$-5$\%$-wi}    &  10  &   3.65 / 0.494 / 2.05 & 0.0211 / 0.0065 / 0.0103 & 35 / 6 / 14 & 135\tabularnewline
{\rm Run-256-100$\rm M_{\odot}$-7$\%$-wio}   &  13  &   2.24 / 0.079 / 0.88 & 0.0224 / 0.0016 / 0.0087 & 33 / 8 / 14 & 182\tabularnewline
{\rm Run-256-100$\rm M_{\odot}$-7$\%$-wi}    &  14  &    6.2 / 0.067 / 1.79 & 0.0211 / 0.0015 / 0.0110 & 36 / 5 / 13 & 175\tabularnewline
{\rm Run-256-100$\rm M_{\odot}$-9$\%$-wio}   &  16  &   2.25 / 0.051 / 0.66 & 0.0239 / 0.0011 / 0.0088 & 74 / 5 / 19 & 304\tabularnewline
{\rm Run-256-100$\rm M_{\odot}$-9$\%$-wi}    &  11  &   1.06 / 0.052 / 0.32 & 0.0231 / 0.0011 / 0.0077 & 23 / 7 / 12 & 134\tabularnewline
        \textbf{Total all models} &  \textbf{70} &   &   &   & \textbf{163}\tabularnewline
                \midrule       
        \midrule 
	\multicolumn{6}{c}{\textbf{2-mag cutoff}}\tabularnewline
	\midrule 
{\rm Run-256-100$\rm M_{\odot}$-5$\%$-wio}   & 3 &  4.20 / 2.136 / 3.17  & 0.0295 / 0.0220 / 0.0265 & 14 / 5 / 10 & 30\tabularnewline
{\rm Run-256-100$\rm M_{\odot}$-5$\%$-wi}    & 7 & 17.03 / 1.338 / 7.02  & 0.0585 / 0.0245 / 0.0366 & 18 / 4 / 10 & 70\tabularnewline
{\rm Run-256-100$\rm M_{\odot}$-7$\%$-wio}   & 3 &  3.01 / 2.014 / 2.40  & 0.0523 / 0.0286 / 0.0414 &  9 / 4 / 8  & 23\tabularnewline
{\rm Run-256-100$\rm M_{\odot}$-7$\%$-wi}    & 3 &  9.13 / 1.248 / 5.84  & 0.0309 / 0.0262 / 0.0285 & 42 / 4 / 25 & 74\tabularnewline
{\rm Run-256-100$\rm M_{\odot}$-9$\%$-wio}   & 5 &  2.99 / 0.207 / 1.57  & 0.0531 / 0.0041 / 0.0308 & 11 / 5 / 8  & 41\tabularnewline
{\rm Run-256-100$\rm M_{\odot}$-9$\%$-wi}    & 4 &  2.29 / 0.243 / 1.09  & 0.0372 / 0.0049 / 0.0261 & 14 / 5 / 10 & 40\tabularnewline
        \textbf{Total all models} &  \textbf{25} &   &   &   & \textbf{46}\tabularnewline
                \midrule        
        \midrule 
        \multicolumn{6}{c}{\textbf{3-mag cutoff}}\tabularnewline
        \midrule 
{\rm Run-256-100$\rm M_{\odot}$-5$\%$-wio}   & 3 & 34.28 / 8.977  / 21.32 & 0.1207 / 0.0910 / 0.1083 & 12 / 3 / 6  & 18\tabularnewline
{\rm Run-256-100$\rm M_{\odot}$-5$\%$-wi}    & 1 & 21.66 / 21.661 / 21.66 & 0.0723 / 0.0723 / 0.0723 & 3  / 3 / 3  & 3\tabularnewline
{\rm Run-256-100$\rm M_{\odot}$-7$\%$-wio}   & 4 & 22.17 / 1.344 / 11.59  & 0.0888 / 0.0286 / 0.0610 & 15/4/7      & 28\tabularnewline
{\rm Run-256-100$\rm M_{\odot}$-7$\%$-wi}    & - & -                      &                        - &      -      &  -\tabularnewline
{\rm Run-256-100$\rm M_{\odot}$-9$\%$-wio}   & - & -                      &                        - & -           & -\tabularnewline
{\rm Run-256-100$\rm M_{\odot}$-9$\%$-wi}    & 2 & 0.83 / 0.608 / 0.72    & 0.0188 / 0.0122 / 0.0155 & 13 / 8 / 11 & 22\tabularnewline
        \textbf{Total all models} &  \textbf{10} &   &   &   & \textbf{18}\tabularnewline
                 \midrule         
        \midrule 
        \multicolumn{6}{c}{\textbf{4-mag cutoff}}\tabularnewline
        \midrule 
{\rm Run-256-100$\rm M_{\odot}$-5$\%$-wio}   & 1 & 23.80 / 23.801 / 23.80 & 0.2313 / 0.2313 / 0.2313 & 2 / 2 / 2 & 2\tabularnewline
{\rm Run-256-100$\rm M_{\odot}$-5$\%$-wi}    & - & - & - & - & -\tabularnewline
{\rm Run-256-100$\rm M_{\odot}$-7$\%$-wio}   & - & - & - & - & -\tabularnewline
{\rm Run-256-100$\rm M_{\odot}$-7$\%$-wi}    & - & - & - & - & -\tabularnewline
{\rm Run-256-100$\rm M_{\odot}$-9$\%$-wio}   & - & - & - & - & -\tabularnewline
{\rm Run-256-100$\rm M_{\odot}$-9$\%$-wi}    & - & - & - & - & -\tabularnewline
        \textbf{Total all models} &  \textbf{1} &   &   &   & \textbf{2}\tabularnewline
                  \midrule       
        \bottomrule
        \hline    
        \end{tabular}
\label{tab:A}
\end{table*}

\begin{figure*}
        \centering
        \includegraphics[width=0.99\textwidth]{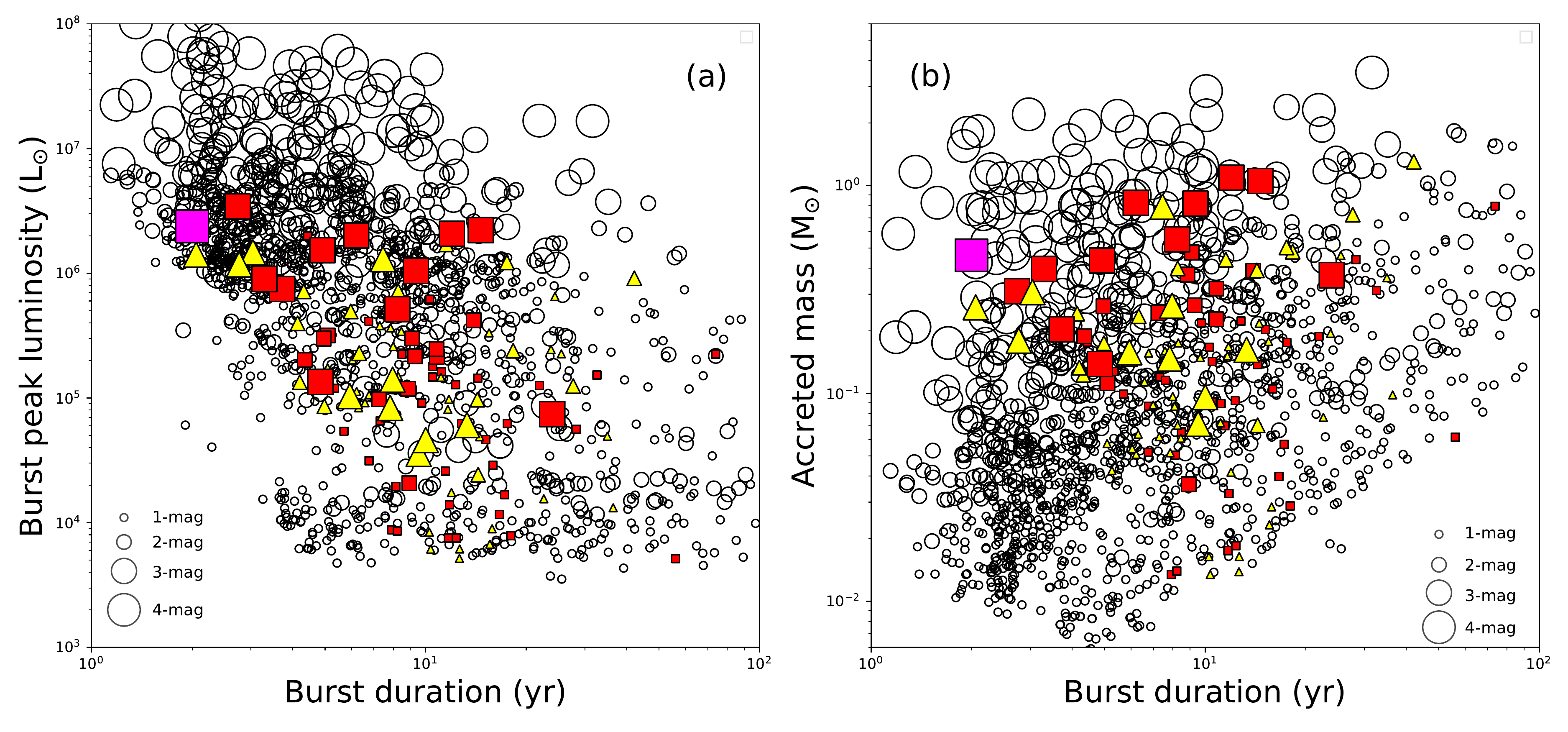}         
        \caption{
        Distribution of the accretion-driven bursts in our series of simulation 
        models in the burst duration-burst peak luminosity (a) and burst duration-burst 
        accreted mass per burst (b) planes. 
        The burst duration is in $\rm yr$, the burst peak luminosity in $\rm L_{\odot}$
        and the accreted mass per burst in $\rm M_{\odot}$, respectively. 
        The burst are displayed for both the disc modelled without (red squares) and 
        with (yellow triangles) stellar inertia. The figure indicates the burst modelled 
        in the previous papers of this series as black circles~\citep{meyer_mnras_500_2021}. 
        The size of the symbols scale with the magnitude of the accretion-driven burst, 
        from the smallest (1-mag bursts), to the largest (4-mag) bursts. 
        \textcolor{black}{Note that the single 4-mag burst of this high-resolution series of 
        models is marked in magenta. }
        \dm{Note that all except one single 4-mag bursts belong to the 
        data in~\citet{meyer_mnras_500_2021}.}
        }
        \label{fig:bursts}  
\end{figure*}

\subsubsection{Disc thermodynamics}
\label{sect:thermodynamics}

In Fig.~\ref{fig:diagram} the time-evolution of the disc midplane \dm{temperature} 
is displayed \dm{as a function of the midplane gas density} for the models with $\beta=7\%$ 
without (left panels) and with wobbling (right panels), for times spanning from 
$17\, \rm kyr$ to $36\, \rm kyr$. Color-coding shows the distance to the 
central protostar $r/R_{\rm disc}$ for each point in the figure, normalised to the disc 
size, with $R_{\rm disc}$ the maximum disc radius at a considered time, which we measure 
by tracking the region at which the infalling material lands onto the disc, using the azimuthaly-averaged 
midplane velocity profiles. The black 
line represents a \dm{fit} to the whole dataset and the upper grayed 
region ($T\ge2000\, \rm K$) \dm{highlights the temperature above which 
molecular hydrogen dissociates. 
%
%
The diagonal branch along which both density and temperature increase as we go from 
larger to smaller distances, as typically protoplanetary discs do. 
The outer diluted and cold disc region (blue points) exhibit less horizontal 
deviation from the diagonal black line than the dense, hot inner part (red 
points), especially at after the beginning of the disc formation 
(Fig.~\ref{fig:discs}-2a, Fig.~\ref{fig:discs}-2e). 
There is a turn-over and a horizontal branch corresponding to the very inner disc 
regions that is warmer than the diagonal region, although density can vary in a wide range. 
Only in this region the gas temperature exceeds the molecular hydrogen dissociation 
threshold ($T \ge 2000\, \rm K$).
}

\dm{
At later times, when the disc has begun to fragment, 
the horizontal deviations in the \dm{inner} region is more pronounced as a result 
of the growth of the disc and scatter over a wider density range. 
The disc structure adopts a more complex morphology provoking additional branches 
in the high temperature part of the diagram, accounting for the many spiral arms 
forming by gravitational instability and enrolling around the sink cell 
(Fig.~\ref{fig:diagram}b). 
At time $29.5\, \rm kyr$ the disc strongly fragments (Fig.~\ref{fig:diagram}c). 
The gaseous clumps extend as a horizontal high temperature branch $\sim 500\, \rm au$ 
from the star. This upper region further evolves as the disc keeps on fragmenting, 
and more clumps form \dm{within} the extended spiral arms, which translates into a 
scattering of the low-temperature region (blue dots of Fig.~\ref{fig:diagram}d). 
}

The disc model with wobbling \DM{included} is qualitatively similar 
to its fixed-star counterpart at time $17\, \rm kyr$ 
(Fig.~\ref{fig:diagram}a,e) in both structure and thermodynamical 
properties, except regarding the disc without stellar inertia is $25\%$ larger since 
its radius reaches $400\, \rm au$ at that time whereas the other disc model 
extends up to $300\, \rm au$ only. 
At time $20\, \rm kyr$ the disc with wobbling does not exhibit the group of 
high temperature and high density point scattering close to the dissociation 
temperature regime (Fig.~\ref{fig:diagram}b,f), because the disc model with 
fixed protostar fragments more violently. 
The absence of early clumps in the model with wobbling results in a distribution 
of the disc midplane gas along the diagonal with much less scattering with 
respect to the black line (Fig.~\ref{fig:diagram}c,g), whereas the inner disc 
region is hotter and exceed the molecular hydrogen dissociation temperature. 
The disc, having fragments located in the inner region ($\le 200\, \rm au$),  
is on the road to multiplicity~\citep{meyer_mnras_473_2018}. 
\dm{
Indeed, disc gas exceeding $2000\, \rm K$ implies second Larson 
collapse~\citep{larson_mnras_145_1969,larson_mnras_157_1972} 
and formation of a protostar, which is not resolved in our simulations, and,  
therefore, multiplicity can only be inferred from this diagram. 
}
At this time, the disc without stellar inertia is more compact than that with wobbling. 
Finally, at time $36\, \rm kyr$ both discs are violently fragmenting and have the 
same overall size of $\approx 1000\, \rm au$. 

\subsection{Bursts}
\label{sect:sub_bursts}

We perform an analysis of the variability in the accretion rate histories onto the young 
massive stellar objects. 
\dm{
The stellar mass and accretion rate measured in the radiation-hydrodynamics simulations 
are turned into a lightcurve via the estimate of the protostellar radius, interpolated 
from the evolutionary tracks of~\citet{hosokawa_apj_691_2009}. 
}
Then, the total luminosity of the central massive young stellar objects is calculated 
for each model as $L_{\rm tot}=L_{\star}+L_{\rm acc}$, with $L_{\star}=L_{\star}(M_{\star},\dot{M})$ 
the photospheric luminosity \dm{from}~\citet{hosokawa_apj_691_2009} and 
$L_{\rm acc} = f G M_{\star} \dot{M} / R_{\star}$ the accretion luminosity, where $G$ 
is the universal gravitational constant, $f=3/4$ a factor accounting for the effective 
mass accreted onto the stellar surface, the rest being released into the a protostellar jet. 
Hence, the variability of the accretion \DM{rate reflects onto evolution} of the total 
luminosity, see fig.~3. of~\citet{meyer_mnras_500_2021}. 
For each model, the lightcurves $L_{\rm tot}(t)$ are post-processed with the signal 
analysis method developed in~\citet{vorobyov_aa_613_2018}, \dm{which}  
filters out the accretion bursts from a synthetic \dm{quiescent} background luminosity 
$L_{\rm bg}$. \dm{During} each burst, the luminosity peak is compared to $L_{\rm bg}$ 
and the ongoing \dm{flare is} classified as an $i$-magnitude ($i$-mag) burst, with $i$ 
such that $L_{\rm tot}\ge2.5^{i}L_{\rm bg}$. 
The method ensures that small variabilities of magnitude $\le i=1$, either from the vicinity 
of the sink cell or from the secular evolution of $L_{\rm bg}$, are not confused with 
\dm{true bursts}~\citep{elbakyan_mnras_484_2019}. The results for the bursts analysis 
are summarised in Table~\ref{tab:A}.

\DM{
Models for the circumstellar medium of young massive stars show that the total number 
of bursts measured in the simulations decreases as their magnitude increases. 
In other words, there are more 1- and 2-mag bursts than 3- and 4-mag flares, and 
a similar tendency is found for  
the total time spent in the burst mode as a function of the burst magnitude, e.g. we 
count 70 1-mag bursts in total and only a single 4-mag burst in our entire series of simulations. 
This trend of declining burst activity of young massive stars is not modified 
when including wobbling in the simulations. 
}
\DM{ 
The effect of disc wobbling on the burst properties can be summarised as follows. 
The number of 1-mag bursts varies at most by up to a factor of $\approx 2$ when 
allowing the protostar to move.  
The number of 2-mag bursts only differs by 1$-$2 bursts in the models with and without wobbling. 
Major differences appear with the 4-mag bursts, i.e. only a model without 
wobbling is able to produce such event, at least for the time interval that we simulate. 
}

\dm{Additionally,} 
higher-magnitude bursts are also much less common than their 1-mag counterparts. 
Only the model with $\beta=5\%$ without wobbling exhibit a short 4-mag burst of duration 
$2\, \rm yr$, suggesting that the realistic discs with stellar inertia do not undergo 
such mechanisms, at least during the early $\approx 40\, \rm kyr$ of their evolution. 
Higher-resolution simulations are required to further investigate the question of 
FU-Orionis-like bursts in massive star formation. 
The average luminosity of the 3-mag bursts is similar regardless of the physics included 
into the models, however, the models without wobbling can accreted slightly more mass per 
bursts.

\begin{table*}
	\centering
	\caption{
	Proportion (in $\%$) of \DM{mass gain} at the end of our simulation, as a function of 
	the flare magnitude during the burst phase of accretion.  
	The intensity of each individual burst is defined as in~\citet{meyer_mnras_482_2019}. 
	}
        \begin{tabular}{lccccccr}
        \hline  
	${\rm {Models}}$  &  $\beta$-ratio  &    ${\rm {Inertia}}$    &   $\rm L_{\rm tot} \sim\, \rm L_{\rm bg}$  &  $1$$-$$\rm mag$ & $2$$-$$\rm mag$ & $3$$-$$\rm mag$ & $4$$-$$\rm mag$ \\ 
        \hline 	
\midrule 
{\rm Run-256-100$\rm M_{\odot}$-5$\%$-wio}    &  $5\, \%$ &  no   &    81.67  &  11.86         &  5.07         &  1.37        &  0.03\tabularnewline
{\rm Run-256-100$\rm M_{\odot}$-7$\%$-wio}    &  $7\, \%$ &  no   &    81.76  &  12.59         &  $\,\,$5.62   &  0.03   &  $\,\,$0.0\tabularnewline
{\rm Run-256-100$\rm M_{\odot}$-9$\%$-wio}    &  $9\, \%$&  no    &    88.10  &  9.02         &  2.88          &  $\,\,$0.0   &  $\,\,$0.0\tabularnewline
\textbf{Mean} & - & - & \textbf{83.84} & \textbf{11.16}  & \textbf{4.53}  & \textbf{0.46}   & \textbf{0.01}\tabularnewline
\midrule 
{\rm Run-256-100$\rm M_{\odot}$-5$\%$-wi}     &  $5\, \%$ &  yes  &    77.54  &  11.17         &  4.61        &  6.65        &  $\,\,$0.0\tabularnewline
{\rm Run-256-100$\rm M_{\odot}$-7$\%$-wi}     &  $7\, \%$ &  yes  &    73.26  &  16.62         & 10.12        &  $\,\,$0.0   &  $\,\,$0.0\tabularnewline
{\rm Run-256-100$\rm M_{\odot}$-9$\%$-wi}     &  $9\, \%$ &  yes  &    84.20  &  10.98         &  4.18        &  0.64   &   $\,\,$0.0\tabularnewline
\textbf{Mean} & - & - & \textbf{78.34} & \textbf{12.92}  & \textbf{6.31}  & \textbf{2.43}  & \textbf{0.0}\tabularnewline
\midrule 
        \hline    
        \end{tabular}
\label{tab:2}
\end{table*}

In Fig.~\ref{fig:bursts} \dm{the burst duration is plotted} as a function of the maximum luminosity 
of each accretion-driven bursts (panel a), and as a function of the 
mass accreted throughout the burst (panel b). On the \dm{panels}, the circles represent 
the data \dm{reported in the previous papers of this 
series which were computed at a lower spatial resolution}~\citep{meyer_mnras_500_2021}, and 
the circle size scales with the burst magnitude. 
The data for the current study are plotted as red squares (models without wobbling) 
and yellow triangles (models with wobbling). 
The bursts of lower magnitude (1-mag) are located in the \dm{$\le 5\times10^{5}\, \rm L_{\odot}$}  
region of the plot, regardless of their duration, spanning from a few $1\, \rm yr$ 
to $\approx 10^{2}\, \rm yr$ (Fig.~\ref{fig:bursts}a). The high magnitude bursts 
(3- and \textcolor{black}{4-mag} bursts) are found in the rather short duration ($\le 20\, \rm yr$) 
and higher luminosity \dm{($> 10^{5}\, \rm L_{\odot}$)} region of the figure, 
although a few bursts without wobbling do not follow this rule and \dm{are located} in 
the long duration region of the figure. 
This is in accordance with simulations of lower spatial resolution (empty circles), 
however, the burst distribution in our simulations do not extend to the same  
upper limit in terms of \dm{the peak luminosity} because of the absence of 4-mag bursts in our 
sample. The bursts are shorter and dimmer compared to a $128\times11\times128$ simulation. 
\dm{
Importantly, the duration of the high-magnitude bursts in the model with wobbling is of 
slightly lower duration than those found in the simulation with fixed star. 
}
In the \dm{burst duration-accreted mass} plane (Fig.~\ref{fig:bursts}b) we see that lower-magnitude bursts 
distribute along diagonals which indicate that the accreted mass scales linearly 
with the burst duration. Once again, as detailed in~\citet{meyer_mnras_500_2021}, 
some of the most mass-accreting bursts are \dm{amongst} the faintest, since the total 
luminosity $L_{\rm tot}$ is dominated by the accretion luminosity $L_{\rm acc} 
\propto 1 / R_{\star}$, with $R_{\star}$ \dm{increasing} during the burst phase, when 
the protostar experiences excursions to the cold part of the Herzsprung-Russell 
diagram~\citep{meyer_mnras_484_2019}.

In Table~\ref{tab:2} we report the proportion of final mass accreted by our protostars 
during the quiescent and the \dm{burst phases}, separated, for the latter, by 
the magnitude of the accretion-driven bursts. 
The mean proportion of final mass accreted during the \dm{burst phase of accretion amounts 
$\approx 17\%$ for the models without wobbling and $\approx 22\%$ when stellar 
inertia is included}, respectively. Similarly, the average mass accreted during 1-mag 
bursts is $\approx 11$$-$$12\%$ {with or without stellar wobbling}, with a slight increase 
as a function of $\beta$-ratio. Differences mostly happen for the 2-mag bursts, with 
$\approx 4.5\%$ and $\approx 6.3\%$ of the total accreted mass with and without 
wobbling, respectively. Similarly, the protostars modelled with stellar inertia gain 
$\approx 2.4\%$ of their mass experiencing 3-mag bursts. None of the models 
accrete mass via 4-mag bursts, since they are absent from the present simulations, 
\dm{except in the model without wobbling and $\beta=5\%$ which produced a single 
occurrence of such a burst}. 
In~\citet{meyer_mnras_482_2019,meyer_mnras_500_2021}, we concluded that massive stars 
gain up to half of their zero-age-main-sequence mass by mass accretion in the burst 
mode, on the basis of models run up to $\sim 70\, \rm kyr$. It is the later phase of 
the disc evolution that the strongest bursts appeared, while we stop our models sooner 
\dm{in the present study, because of the enormous computational costs of our 
higher-resolution calculations}.


\begin{figure}
        \centering
        \includegraphics[width=0.425\textwidth]{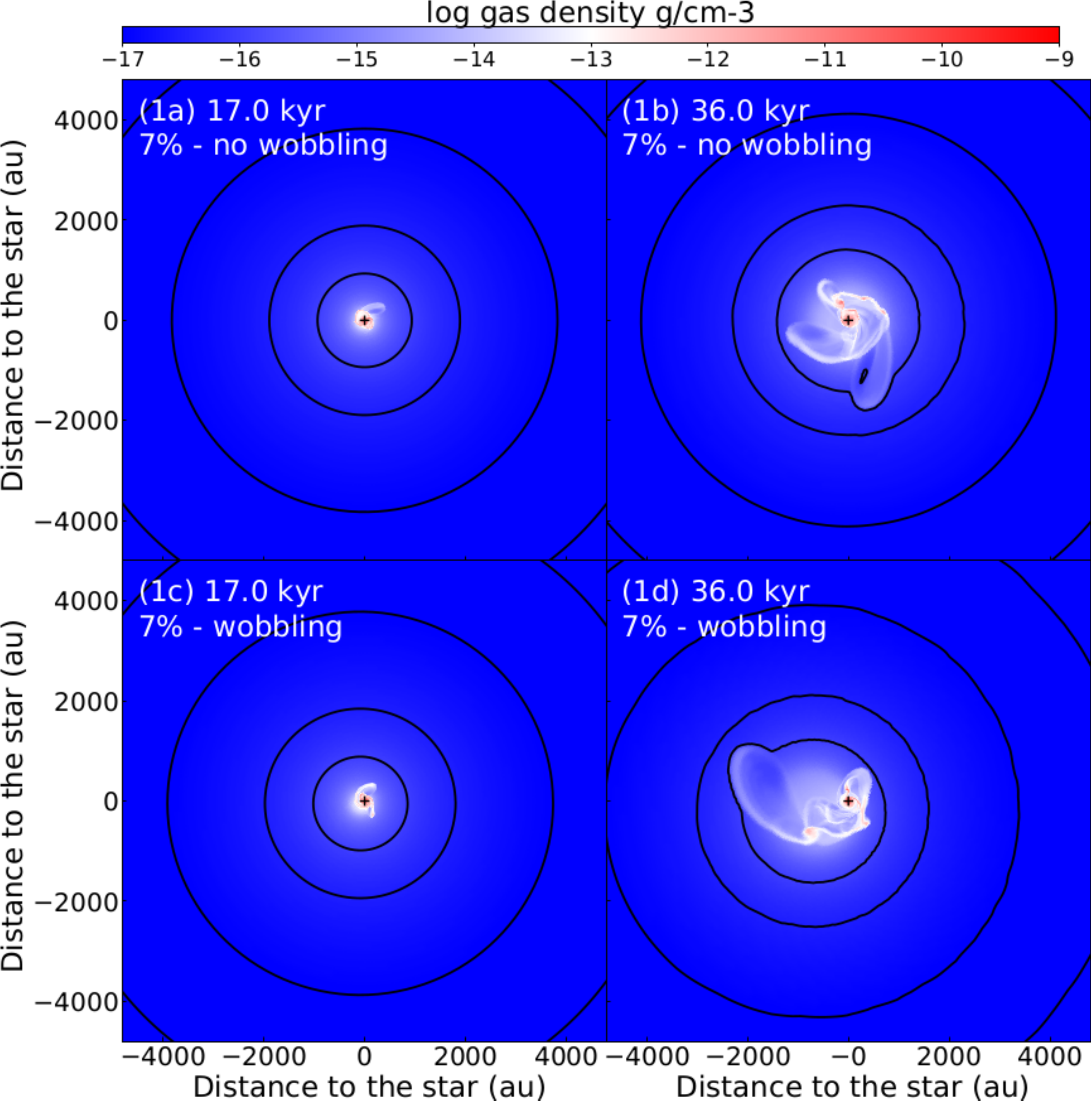} \\
        \includegraphics[width=0.425\textwidth]{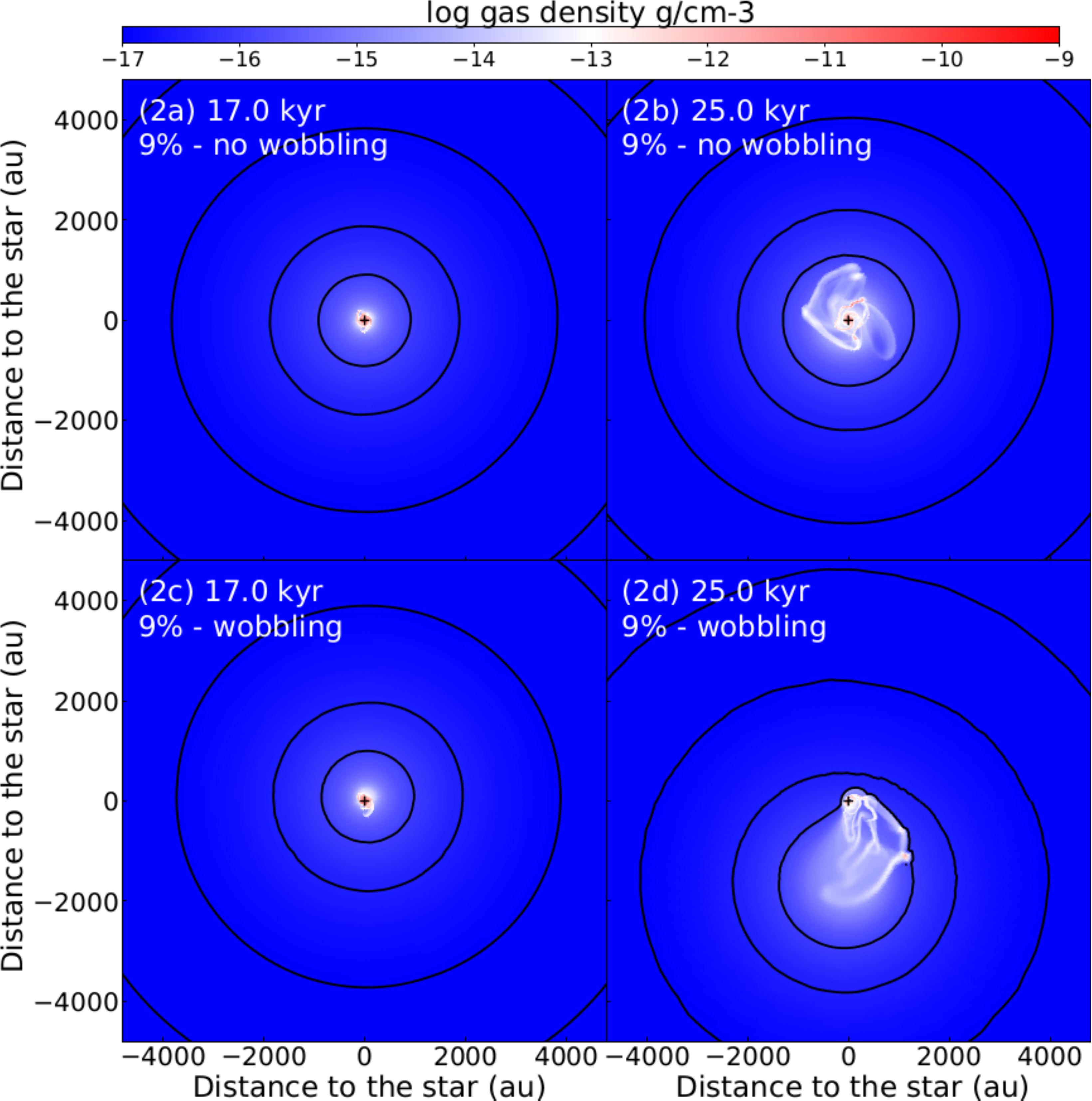}   
        \caption{
        Disc midplane with $\beta=7\%$ (upper series of panels) and 
        $\beta=9\%$ (lower series of panels), and, each of them displaying both models with 
        (top) and without (bottom) wobbling. 
        The figures show the disc midplane \textcolor{black}{gas 
        density, (in $\rm g\, \rm cm^{-3}$)} for several selected characteristics 
        time instances of the accretion disc evolution, and with solid black number 
        density isocontours highlighting the effects of the stellar inertia 
        onto the mass distribution in the computational domain. 
        }
        \label{fig:wobbling}  
\end{figure}

\section{Discussion}
\label{sect:discussion}

This section reminds the reader the limitations of our simulation setup and discuss the observability 
of the accretion disc models if observed by means of the {\sc alma} interferometer.

\begin{figure*}
        \centering
        \includegraphics[width=0.75\textwidth]{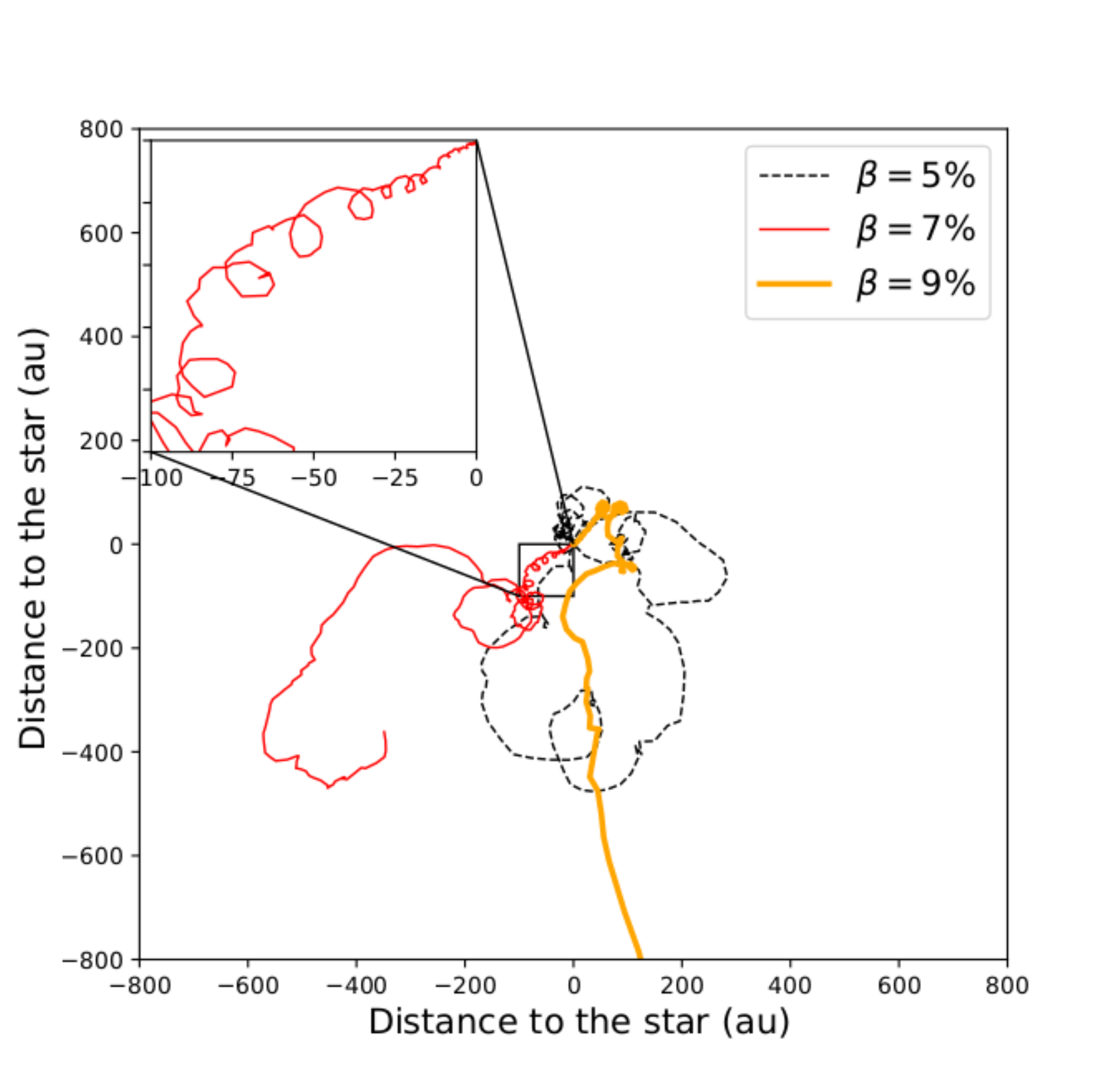}          
        \caption{
        Time evolution of the position of the center of mass of the accretion 
        discs modelled with stellar inertia (in $\rm au$), for our simulation
        with $\beta=5\%$ (black dashed line), $\beta=7\%$ (red thin solid line)
        and $\beta=9\%$ (orange thick solid line). Inset plot zooms on the 
        wobbled motion induced by the stellar inertia. 
        }
        \label{fig:barycenter}  
\end{figure*}

\subsection{Caveats}
\label{sect:discussion_caveats}

Most of the limitations in the numerical method utilised in this study have been presented 
in the parameter study~\citep{meyer_mnras_500_2021}, particularly regarding the absence of 
magnetic fields, non-ideal magneto-hydrodynamics and photoionisation that mostly 
affect the bipolar jet filled with hot gas. \dm{While} the magnetic feedback should be improved in 
future studies, not taking into account the protostellar ionization is acceptable  
since we concentrate onto the disc structure and its properties. Nevertheless, disc 
evaporation and disc wind are neglected in the simulations. 
This work explores the stellar motion induced by the disc in the fashion 
of~\citet{michael_mnras_406_2010,regaly_aa_601_2017} 
with \dm{moderate}-resolution $256\times41\times256$ simulations, which is a major improvement 
relative to the models presented in~\citet{meyer_mnras_473_2018} 
and~\citet{meyer_mnras_482_2019} since our work is based on both high-resolution models 
performed with stellar inertia. 
Nevertheless, the midplane symmetry that we impose to our calculations intrinsically 
neglects any vertical motion of the massive protostar, although low-resolution tests 
suggested that this effect is not important~\citep{meyer_mnras_473_2018}. In future studies 
we will consider higher-resolution models like those presented in~\citet{oliva_aa_644_2020}.  
The \dm{increased} resolution of our simulations imposed to \dm{move} the size of the inner radius 
of the sink cell $r_{\rm in}$ to $20\, \rm au$ instead of $10\, \rm au$ 
in~\citet{meyer_mnras_473_2018,meyer_mnras_482_2019}. As the time-step restrictions are 
principally governed by the size of the innermost cell of the radial logarithmically 
expanding mesh, we had to keep the calculations affordable from the point of view of the 
computational costs.

\begin{figure*}
        \centering
        \includegraphics[width=0.8\textwidth]{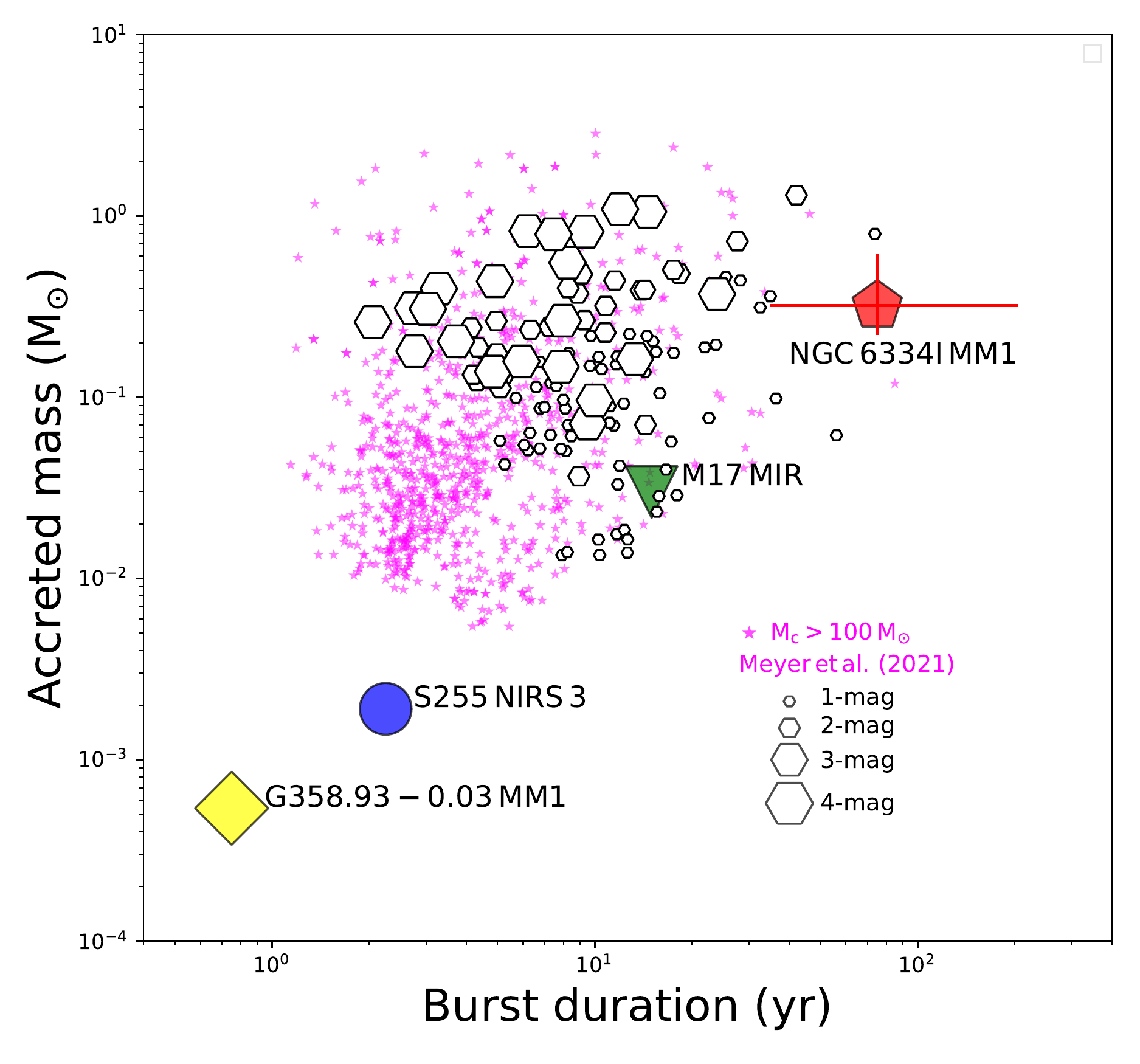}          
        \caption{
        \textcolor{black}{
        Comparison of our modelled bursts with the properties of observed bursts 
        in the burst duration-accreted-mass diagram.
        The white hexagons are the burst from the simulated discs presented in this study, 
        the coloured symbols stand for the burst observed so far from young high-mass stellar 
        objects. The magenta stars represent the burst from the models with an initial 
        molecular core model $M_{\rm c}\ge100\, \rm M_{\odot}$ in~\citet{meyer_mnras_500_2021}. 
        }
        }
        \label{fig:observations_comparison}  
\end{figure*}

As observed in~\citet{michael_mnras_406_2010} and~\citep{regaly_aa_601_2017} in the context of 
accretion discs of low-mass 
stars, stellar inertia delays the development of gravitational instabilities in the disc 
and modifies its overall appearance, in particular the \dm{spatial morphology of the 
structures} growing in it. 
While the consideration of stellar inertia adds \dm{realism} of the simulated models, it \dm{also} 
introduces an additional boundary \dm{effect related to} the displacement of the disc in the computational 
domain that mimics the stellar displacement \dm{in the non-inertial frame of reference}. 
This particularly takes place when 
the mass of the disc is comparable to that of the protostar $M_{\rm d}/M_{\star}\approx 1$, 
as it is the case in the formation of our massive stars (Fig.~\ref{fig:mass_disc}b, 
see also our model with $\beta=7\%$ in Fig.~\ref{fig:discs}-3h). 
However, the timescales over which high-mass stars \DM{operate} ($\sim \rm kyr$) is much shorter 
than that of their lower mass counterparts ($\rm Myr$), and, therefore, the boundary effects 
are somehow less pronounced in our models except in the simulation with high $\beta$-ratio. 
An updated version of the code should handle moving boundaries and/or replenishing the 
void caused \dm{by means of the} displacement of the density field by infalling pre-stellar core material 
that is initially out of the computational domain at radii $r \ge R_{\rm max}$.

\subsection{Effect of wobbling on disc structure}
\label{sect:discussion_emission}

Fig.~\ref{fig:wobbling} shows a zoomed-out view of the midplane density field in 
($\rm g\, \rm cm^{-3}$) of the computational domain in our simulations with 
$\beta=7\%$ (top panels) and $\beta=9\%$ (bottom panels) for several selected 
time instances of the system evolution. 
The figures cover a region of $5000\times5000\, \rm au$ including both the accretion 
disc in the inner $\le 1000\, \rm au$ and the still-collapsing pre-stellar core.  
Density isocontours as solid black lines highlight the structure of the 
density field of the molecular envelope and the disc. 
At $17\, \rm kyr$, the midplane density \dm{in the collapsing envelope is} 
axisymmetric as were the initial conditions for the pre-stellar core 
(Eq.~\ref{eq:density_profile}), in both simulation with and without wobbling, 
see also Fig.~\ref{fig:wobbling}~1a-1c (top). 
Almost at the end of the simulation (at time $\ge 36\, \rm kyr$) the deviation from the 
axisymmetry of the initial density structure is pronounced in the case with stellar 
inertia (Fig.~\ref{fig:wobbling}~1d), as the stellar motion affects not only the 
accretion disc located in the inner region of the computational domain, but also 
the whole envelope. 
For fast initial rotation of the pre-stellar core ($\beta\ge9\%$), the deviations 
from axisymmetry are magnified and happen quicker, at time $25\, \rm kyr$, see  
Fig.~\ref{fig:wobbling}~2d (bottom). The catastrophic effects of stellar wobbling on the 
\dm{envelope and disc} evolution are caused by the fact that  
\dm{the acceleration caused by the non-inertial frame of reference} 
shifts the material in the entire computational domain. 
\dm{
As a result, a distortion in the gas density near the outer fixed boundary 
develops, which gradually feeds back and modifies through gravity the entire 
evolution of the inner structure. 
}

\begin{figure*}
        \centering
        \includegraphics[width=0.99\textwidth]{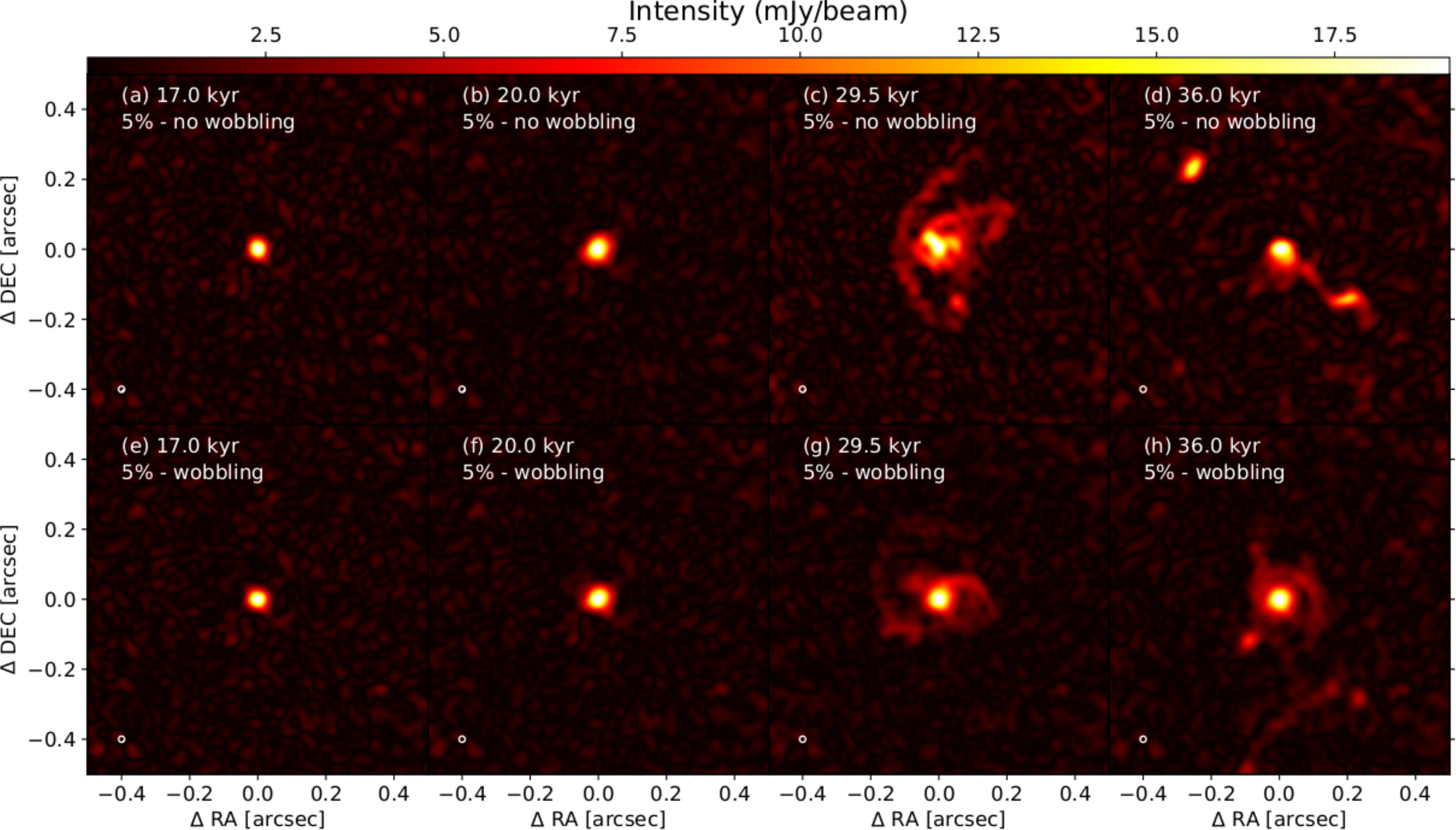} \\
        \caption{
        Synthetic emission maps of the models with $\beta=5\%$, without (top panels) and 
        with (bottom panels) stellar inertia, as seen by the {\sc ALMA} interferometer in 
        its antenna configuration 10. The disc are shown at the same characteristic 
        times as for the upper series panels in Fig.~\ref{fig:discs}. 
        \dm{The distance to the source is assumed to be $1\, \rm kpc$.}
        }
        \label{fig:synth_maps}  
\end{figure*}

We display the evolution of the position of the barycenter of the star-disc-envelope 
system in Fig.~\ref{fig:barycenter}. The center of mass is calculated as, 
\begin{equation}
	  x_\mathrm{bar}(t) = \frac{1}{M} \sum_{\rm ijk} x_{\rm ijk} \rho_{\rm ijk} dV_{\rm ijk}, 
      \label{eq:bar_x}
\end{equation}
\begin{equation}
	  y_\mathrm{bar}(t) = \frac{1}{M} \sum_{\rm ijk} y_{\rm ijk} \rho_{\rm ijk} dV_{\rm ijk},  	  
      \label{eq:bar_y}
\end{equation}
and, 
\begin{equation}
	  z_\mathrm{bar}(t) = \frac{1}{M} \sum_{\rm ijk} z_{\rm ijk} \rho_{\rm ijk} dV_{\rm ijk},  	  
      \label{eq:bar_z}
\end{equation}
where,
\begin{equation}
x_{\rm ijk}=r_{\rm ijk}\cos(\phi_{\rm ijk})\sin(\theta_{\rm ijk})
\end{equation}
\begin{equation}
y_{\rm ijk}=r_{\rm ijk}\cos(\phi_{\rm ijk})\sin(\theta_{\rm ijk})
\end{equation}
and,
\begin{equation}
z_{\rm ijk}=r_{\rm ijk} \cos(\theta_{\rm ijk})
\end{equation}
are the Cartesian coordinates of a disc volume element $ijk$, respectively. 
The midplane symmetry of the computational domain imposes $z_{\rm bar}=0$ 
and the whole mass $M$ included into it reads, 
\begin{equation}
	  M = \iiint \rho dV =  \sum_{\rm ijk} \rho_{\rm ijk} dV_{\rm ijk}, 	  
      \label{eq:mass}
\end{equation}
where $dV_{\rm ijk}$ is the volume element of the grid zone $ijk$ in the computational 
domain, with $1\le i \le 256$, $1\le j \le 41$ and $1\le k \le 256$. 
The barycenter exhibits large excursions from the origin of the computational domain 
($x=0, y=0$, $z=0$) as well as smaller-scale periodic deviations from the path of the excursions. 
The large-scale motion is a direct consequence of the non-axisymmetric shape adopted 
by the infalling molecular material of the envelope and, as the $\beta$-ratio of the 
model increases, the excursions of the star away from the origin are more important 
(see yellow line of Fig.~\ref{fig:barycenter}). 
The small amplitude displacement of the barycenter \dm{is caused by} of the development of 
substructures in the disc, such as dense spiral arms and gaseous clumps, which are 
of much milder effects.

\begin{figure*}
        \centering
        \includegraphics[width=0.95\textwidth]{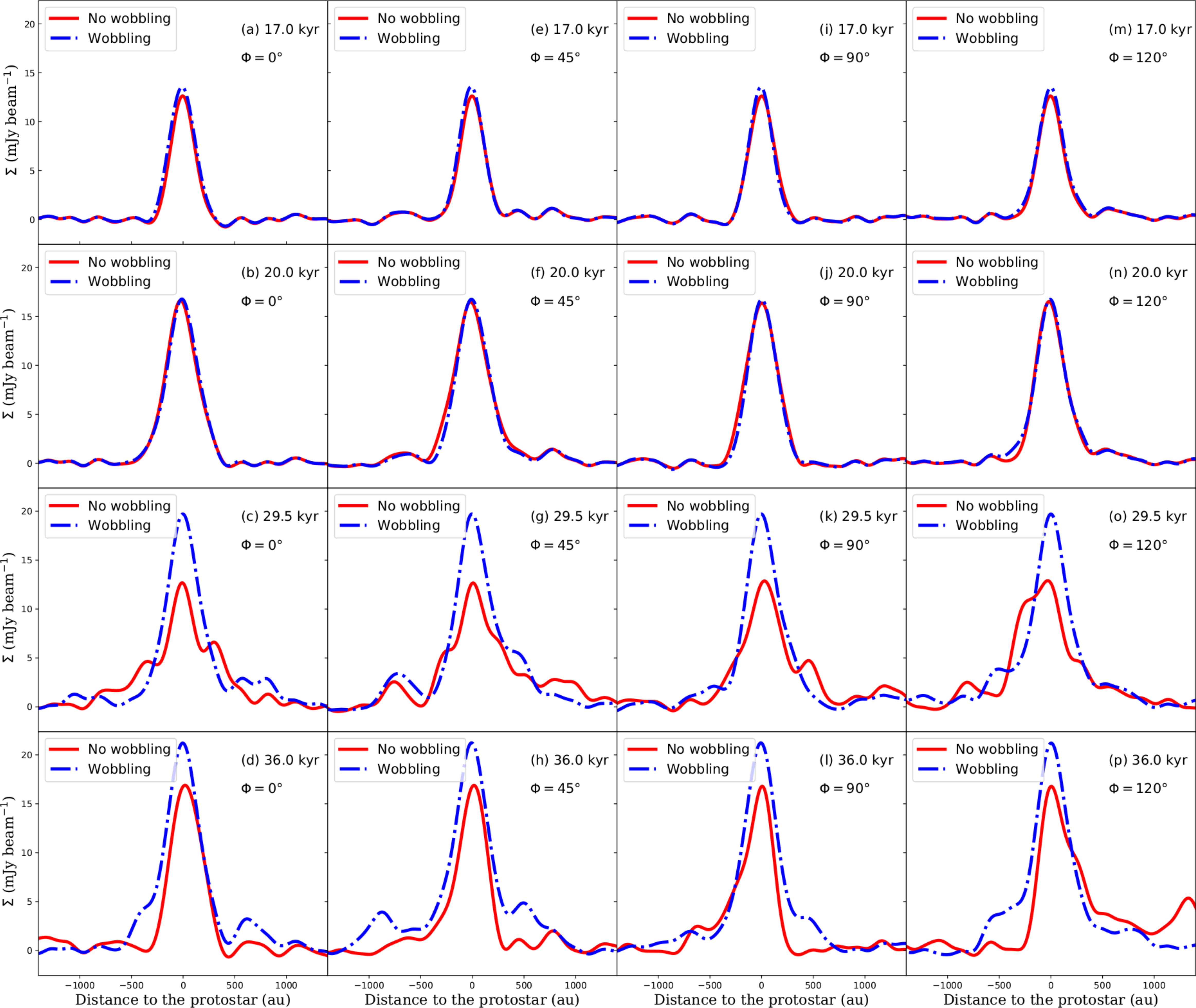} \\
        \caption{
        Cuts through the synthetic emission maps of our accretion disc model 
        with $\beta=5\%$ (in $\rm mJy\, \rm beam^{-1}$) for the inner disc region. 
        Each panel show the surface brightness without (solid red line) and with 
        stellar inertia (dashed blue line). The horizontal lines of panels display 
        the data for a same time instance, and each column correspond to a different 
        cut through the emission maps, measured clock-wise, and characterised by the 
        angle $\phi$ of the cut with respect to its vertical north-south direction. 
        }
        \label{fig:synth_maps_cuts}  
\end{figure*}

\subsection{\dm{Comparison with monitored bursts}}
\label{sect:comparison_observation}

\textcolor{black}{
Fig.~\ref{fig:observations_comparison} compares our modelled bursts with the properties of 
observed flares, in the burst duration-accreted mass diagram. The white hexagons are 
the burst measures from the lightcurves calculated on the basis of the accretion rate histories 
of our simulated protostars and the coloured symbols stand for real bursts of high-mass stars. 
\dm{
The red pentagon represents the burst of NGC~6334I~MM1, which duration has 
been constrained to $t_{\rm bst}\approx 40$$-$$130\, \rm yr$~\citep{hunter_apj_912_2021} 
and the correspondingly accreted mass estimated to $M_{\rm bst}\approx0.1$$-$$0.3\, 
\rm M_{\odot}$~\citep{hunter_apj_912_2021,elbakyan_aa_651_2021}. 
}
Similarly, the yellow diamond marks the burst of G358.93-0.003~MM1, 
which properties are $t_{\rm bst}\approx 0.75\, \rm yr$ and $M_{\rm bst}\approx0.566\, 
\rm M_{\rm J}$~\citep{stecklum_aa_646_2021,elbakyan_aa_651_2021}.  
The blue circle and the green triangle stand for the bursts monitored from the 
young massive stellar object S255~NIRS~3 ($t_{\rm bst}\approx 2.25\, \rm yr$,
$M_{\rm bst}\approx2\, \rm M_{\rm J}$), see~\citet{caratti_nature_2016,elbakyan_aa_651_2021}, 
and that of the recently-discovered outbursts of M17~MIR ($t_{\rm bst}\approx 15\, \rm yr$,
$M_{\rm bst}\approx31.43\, \rm M_{\rm J}$), see~\citet{2021arXiv210812554C,elbakyan_aa_651_2021} 
and references therein. 
}

\textcolor{black}{
The bursts observed from S255~NIRS~3 and G358.93-0.003~MM1 have a smaller 
amount of accreted mass compared to that found in our simulations, while their duration 
is consistent with our data, although G358.93-0.003~MM1 is shorter by an order 
of magnitude than the shortest burst duration measured from our models. 
This, however, can be explained by our too big inner sink cell radii. The use of a
smaller sink would give the infalling gaseous clumps more time to lose their 
outer envelope via tidal stripping when migrating closer to the star, as 
studied in the so-called tidal downsizing model~\citep{nayakshin_mnras_454_2015,nayakshin_2015,nayakshin_mnras_452_2015,
nayakshin_mnras_461_2016}. Consequently, the actual mass of the clump would 
become smaller as it approaches the star and only the core would survive. 
\dm{
The accreted mass value observed for the burst of NGC~6334I~MM1 is consistent with 
our modelled burst. This burst being classified as a 3-mag burst in the parlance 
of our simulations for the burst mode of accretion in massive star 
formation~\citep{hunter_apj_912_2021}, however, our models predict that such burst 
should be of slightly shorter duration. Such discrepancy could be explained by the fact 
that the initial properties of the molecular cloud in which NGC~6334I~MM1 forms might
probably be different than that assumed in our simulations. 
}
The outburst of M17~MIR shows much more consistencies with our predictions, both in terms 
of duration and accreted mass (green triangle). 
%
%
The magenta stars of Fig.~\ref{fig:observations_comparison} are the bursts from the models 
with initial molecular cores $M_{\rm c}\ge 100\, \rm M_{\odot}$~\citep{meyer_mnras_500_2021}, 
which have smaller accreted mass than the bursts of the present study which all assume 
$M_{\rm c}=100\, \rm M_{\odot}$. We speculate that the initial core of M17~MIR is perhaps 
of smaller mass than that responsible for the formation of the other protostars, i.e. 
S255~NIRS~3 and G358.93-0.003~MM1. Future studies tailored to particular 
bursts should take this into account. 
}

\textcolor{black}{
\textcolor{black}{Last, let's also} underline that the 
observational measures of the burst properties 
are affected by a certain number of uncertainties. 
The burst properties are principally \textcolor{black}{derived by measuring the variability} in 
brightness temperature of the object, assumed to be equal to that of the dust 
temperature, from which the luminosity is calculated. This relies on geometrical 
effects potentially scattering radiation out of the observer's light-of-sight, and 
therefore decreasing the monitored brightness temperature and radiation 
intensity~\citep{hunter_apj_837_2017}. Additionally, absorption effects are at 
work, reducing the observed dust continuum flux and drastically reducing the 
observed accretion luminosity from which the accreted mass is 
derived~\citep{johnston_aa_551_2013}. 
Consequently, the real data in the burst duration-accreted mass diagram, represented 
by coloured symbols in Fig.~\ref{fig:observations_comparison} are not firm numbers, 
but high uncertainties estimates. Future observational work on the finer 
characterisation of burst properties would be highly desirable. 
}

\subsection{\DM{Emission maps and observability}}
\label{sect:discussion_emission}

We repeat the exercise of calculating predictive {\sc alma} images for the simulated 
discs as done in~\citet{meyer_487_MNRAS_2019}, see 
also~\citet{macfarlane_mnras_487_2019,macfarlane_mnras_487_2019b,skliarevskii_arep_65_2021}. 
The dust density field calculated with the {\sc pluto} code, taking into account the 
information about dust-sublimated regions, is imported into the radiative transfert 
code {\sc radmc-3d}\footnote{http://www.ita.uni-heidelberg.de/dullemond/software/radmc-3d/
}~\citep{dullemond_2012} for the characteristic simulation snapshots of the model 
with $\beta=5\%$ in Fig.~\ref{fig:discs}. 
It first calculates the dust temperature by Monte-Carlo 
simulation using the method of~\citet{bjorkman_apj_554_2001} before ray-tracing  
$10^{8}$ photons packages from the stellar surface to the disc. Then, it performs 
radiative transfer calculations against dust opacity assuming that the dust in the 
disc is the silicate mixture of~\citet{laor_apj_402_1993}. The source of photons 
is considered to be a black body of effective temperature $T_{\rm eff}$ determined 
for each disc snapshots using the stellar evolutionary tracks of~\citet{hosokawa_apj_721_2010}.

Synthetic images of the disc are produced for a field of view covering a radius of 
$1000\, \rm au$ around the central protostar and made of $2000\times2000$ grid zones 
onto which the emissivity is projected. Images are calculated at $249.827\, \rm Ghz$ 
($1.2\, \rm mm$) with a channelwidth of $50.0\, \rm Mhz$ which corresponds to the 
{\sc alma} band 6, assuming no inclination angle for the accretion disc with respect to 
the plane of the sky. 
Last, the simulated emission maps are treated with the Common Astronomy Software 
Applications {\sc casa}\footnote{https://casa.nrao.edu/}~\citep{McMullin_aspc_376_2007}.  
We obtain simulated interferometric images which can be directly compared to the data 
acquired by the {\sc alma} facility. The antennae are assumed to be in the 
most extended, long-baseline spatial configuration C43-10 using 43 $12\, \rm m$ 
antennae permitting to reach a maximal spatial resolution of $0.015^{"}$. 
\dm{The distance to the source is assumed to be $1\, \rm kpc$.}
We refer the reader interested in further details on our method for simulating the 
synthetic images~\citep{meyer_487_MNRAS_2019}.

Fig.~\ref{fig:synth_maps} plots the $1.2\, \rm mm$ synthetic {\sc alma} emission maps 
of the disc model with the rotation-to-gravitational ratio $\beta=5\%$, without 
(top panels) and with stellar inertia (bottom panels), for the characteristic time 
instances of the disc evolution presented in Fig.~\ref{fig:discs}-1. 
The images at time $17\, \rm kyr$ consist of a bright circle representing the nascent 
disc, right after the end of the \dm{free-fall gravitational collapse of the cloud 
material} (Fig.~\ref{fig:synth_maps}a). This 
remains unchanged up to time $20\, \rm kyr$, since the star has not yet entered the burst 
mode and keeps on gaining mass by quiescent accretion from a stable and unfragmented 
disc (Fig.~\ref{fig:discs}-1b). The model with stellar inertia \dm{does} not exhibit identifiable 
substructures either at a similar time of evolution (Fig.~\ref{fig:discs}-1e-f). 
At time $29.5\, \rm kyr$ the circumstellar medium of the massive protostar is made 
of a bright central stellar region, accompanied by a visible infalling gaseous clump in 
the close stellar surroundings. The entire structure is surrounded by a large-scale 
spiral arm wrapped around the inner disc (Fig.~\ref{fig:synth_maps}c). Note that the 
very extension of the spiral arm is not visible, hence, the observed disc does not 
fully reflect the entire size and composition of the whole object, see midplane 
density field in  Fig.~\ref{fig:discs}-1c. 
Fragmentation further modifies the disc millimeter appearance which takes the typical 
morphology of a massive double protosystem, see image Fig.~\ref{fig:synth_maps}d.

The simulation model including stellar inertia displays a typical two-armed structure 
at $29.5\, \rm kyr$ (Fig.~\ref{fig:synth_maps}g) and a more complex pattern at the end of 
the simulation, including an inner enrolled spiral arms, gaseous clumps and more extended 
arms with fainter clumps in the southern region (Fig.~\ref{fig:synth_maps}h). 
\textcolor{black}{The} spiral arms reveal 
a certain level of granulosity arising from the interferometric 
measures, not originating from the disc structure itself (Fig.~\ref{fig:synth_maps}c), 
and this can be misinterpreted as additional disc fragments as of the same brightness 
as the faintest gaseous clumps. 
These models take their importance in the context of the recently observed bursting young 
massive protostars and their circumstellar 
environments~\citep{johnston_apj_813_2015,chen_apj_823_2016,ilee_mnras_462_2016,
forgan_mnras_463_2016,maud_467_mnras_2017,chen_apj_835_2017,maud_aa_620_2018,ahmadi_aa_618_2018,
2018ApJ...869L..24I,maud_aa_620_2018,sanna_aa_623_2019,motogi_apj_877_2019,maud_aa_627_2019,
liu_apj_904_2020,sanna_aa_655_2021,suri_aa_655_2021,humphries_mnras_502_2021,Vorster_2021,
williams_mnras_509_2022,mccarthy_mnras_509_2022}.

Fig.~\ref{fig:synth_maps_cuts} displays cross-sections through the simulated images of 
the disc models without (solid red line) and with (dashed blue line) stellar wobbling plotted in 
Fig.~\ref{fig:synth_maps} for several angles $\phi=0\degree$, $45\degree$, $90\degree$ 
and $\phi=120\degree$ with respect to the north-south axis. 
No notable differences happen at time $17.0\, \rm kyr$ between the simulation models 
without and with stellar inertia (Fig.~\ref{fig:synth_maps_cuts}a,e,i,m). This persists 
up to time $20.0\, \rm kyr$ in the sense that the two accretion discs models remain 
undistinguishable in terms of emission properties (Fig.~\ref{fig:synth_maps_cuts}b,f,j,n). 
The differences begin to be evident at times $\ge 29.5\, \rm kyr$. The models with 
stellar inertia are brighter at the center of the disc simulated without wobbling, 
see blue line in Fig.~\ref{fig:synth_maps_cuts}c. However, the discs \dm{without 
stellar wobbling} display higher surface brightnesses in the inner $\le 800\, \rm au$ 
region, see Fig.~\ref{fig:synth_maps_cuts}c,g,k,o, whereas this trend 
is inverted at later time ($36.0\, \rm kyr$), i.e. the disc with stellar inertia 
emits more \dm{than} that modelled without wobbling, see 
Fig.~\ref{fig:synth_maps_cuts}d,h,l,p. 
\DM{
Consequently, the disc model without wobbling 
underestimates the observability of the circumstellar medium of massive stars  
and produces massive binary/multiple systems easily (Fig.~\ref{fig:synth_maps_cuts}p). 
}
It is therefore necessary to include stellar wobbling in high-resolution simulations 
in order to reduce this over-fragmentation generated by the simulations presented  
in~\citet{meyer_487_MNRAS_2019}.

\section{Conclusion}
\label{sect:conclusion}

\DM{
In this paper, we investigate the formation of young high-mass stars by means of high-resolution 
simulations. 
We perform three-dimensional gravito-radiation-hydrodynamical simulations with the 
{\sc pluto} code~\citep{mignone_apj_170_2007,migmone_apjs_198_2012,vaidya_apj_865_2018} 
of several collapsing pre-stellar cores of different kinetic-to-gravitational energy 
ratio $\beta=5\%$$-$$9\%$. 
The performed numerical models are of spatial resolution equivalent to the 
highest resolution model presented in~\citet{meyer_mnras_473_2018}, hence, 
they are better resolved than in the parameter study~\citep{meyer_mnras_500_2021}. 
Stellar motion is included in the simulation models via the introduction 
of an additional gravitational acceleration felt by the disc and envelope in 
the non-inertial frame of reference centered on the 
star~\citep{Michael_2010MNRAS,hosokawa_2015,regaly_aa_601_2017}. 
We analyse the various disc and pre-main-sequence accretion bursts properties, and, 
by means of radiative transfer calculations against dust opacity in the millimeter 
waveband, we explore the effect of the stellar motion onto the observability of the 
disc by the {\sc alma} interferometer.  
}
%

\DM{
The growing non-axisymmetric disc substructures formed by gravitational 
instability such as spiral arms and gaseous clumps~\citep{meyer_mnras_473_2018} exert 
onto the protostar a force sufficient to off-set the barycenter of the star-disc 
system from the geometrical center of the computational domain. 
The morphology of the accretion discs is affected by the inclusion of stellar inertia into 
the simulations. 
\textcolor{black}{
The angular momentum of the disk decreases because part of it is transferred to the star,
and, consequently, the discs fragment later, retain a rounder and more compact shape, 
and, form fewer gaseous clumps. 
}
The discs with wobbling eventually develop spiral arms from which migrating gaseous 
clumps form and undergo the mechanism depicted in~\citet{meyer_mnras_464_2017}. 
Our models with stellar inertia experience \textcolor{black}{fewer} high-magnitude accretion bursts 
for fast initial pre-stellar core rotation ($\beta\ge7\%$), than the models without 
stellar inertia. The properties of these bursts in our models with stellar inertia are 
in good agreements with that monitored from the young high-mass star 
M17~MIR~\citep{2021arXiv210812554C}.
}

Prediction for the interferometric observability of the disc models is  
\dm{calculated} with radiative transfer calculations using the 
code {\sc radmc-3d}~\citep{dullemond_2012} and the synthetic imaging code 
{\sc casa}~\citep{McMullin_aspc_376_2007}, in order to obtain $1.2\, \rm mm$ 
simulated {\sc alma} observations. 
\DM{
Disc fragments are visible for a Cycle 10 {\sc alma} 
observation with the C-43 antenna configuration, as long as 
the disc is old enough to have experienced efficient gravitational fragmentation, 
and that the star entered the burst mode of accretion. 
Circumstellar clumps and other substructures should be searched in the inner 
$\approx 500-600\, \rm au$ of the disc of massive protostars. 
}

The presented models should be improved in the future, both in terms of \dm{realism} 
of the disc by including additional physical mechanisms 
\DM{and accurately treating the boundary effects in the non-inertial frame of reference,}
but also by extending 
the radiative transfer calculations to other wavebands of the electromagnetic 
spectrum in order to investigate the spectral evolution of the circumstellar medium 
of massive protostars~\citep{frost_apj_920_2021}.


\section*{Acknowledgements}

\DM{
The authors thank the anonymous referee for comments which improved the quality of the paper.
}
\dm{
DMA Meyer thanks T.~Hosokawa for technical advice on stellar inertia. 
}
The authors acknowledge the North-German Supercomputing Alliance (HLRN) for providing HPC resources 
that have contributed to the research results reported in this paper. V.G. 
Elbakyan acknowledges support from STFC grants ST/N000757/1 to the University of Leicester. 
E. I. V. and A. M. S. acknowledge support of Ministry of Science and
Higher Education of the Russian Federation under the grant 075-15-2020-780.
S.~Kraus acknowledges support from the European Research Council through an 
Consolidator Grant (Grant Agreement ID 101003096), and an STFC Consolidated 
Grant (ST/V000721/1).
\dm{
SY~Liu acknowledges support from the grant MOST 108-2923-M-001-006-MY3. 
}

\section*{Data availability}

This research made use of the {\sc pluto} code developed at the University of Torino  
by A.~Mignone (http://plutocode.ph.unito.it/). The figures have been produced using the 
Matplotlib plotting library for the Python programming language (https://matplotlib.org/). 
The data underlying this article will be shared on reasonable request to the corresponding 
author.


\bibliographystyle{mn2e}

\footnotesize{
\bibliography{grid}
}


\end{document}